\documentclass[12pt,dvipsnames]{article}
\usepackage[top=80pt,bottom=85pt,left=85pt,right=85pt]{geometry}

\pdfoutput=1
\usepackage[utf8]{inputenc}
\usepackage[T1]{fontenc} 
\usepackage[english]{babel} 
\usepackage{braket}
\usepackage[dvipsnames]{xcolor} 
\usepackage{physics}
\usepackage{tabularx} 
\usepackage{ragged2e}
\usepackage{booktabs} 
\usepackage{rotating}
\usepackage{makecell,booktabs}
\usepackage{multirow} 
\usepackage{comment} 
\usepackage{bbm}
\usepackage{amsmath}
\usepackage{cite}
\usepackage[all]{xy}

\usepackage{tikz}
\usetikzlibrary{math} 
\usetikzlibrary{3d} 
\usepackage{tikz-3dplot}
\usepackage{tikz-cd} 
\usepackage{microtype}
\usepackage{amsthm}
\usepackage{mathrsfs} 
\usepackage[strict]{changepage}
\newcommand\scalemath[2]{\scalebox{#1}{\mbox{\ensuremath{\displaystyle #2}}}}

\tikzcdset{
  cells={font=\everymath\expandafter{\the\everymath\displaystyle}},
}

\usepackage{amssymb} 
\usepackage{subcaption} 
\DeclareCaptionFormat{custom}

\usepackage[pdftex,colorlinks=true]{hyperref} 
\hypersetup{urlcolor=MidnightBlue, citecolor=red, linkcolor=MidnightBlue}

\usepackage{float} 
\usepackage{todonotes}

\usepackage[capitalise]{cleveref}

\newcolumntype{E}{>{\hfil$}p{0.65cm}<{$\hfil}}

\newcolumntype{L}{>{\hfil$}p{16cm}<{$\hfil}}

\newcolumntype{D}{>{\hfil$}p{7.4cm}<{$\hfil}}
\newcolumntype{C}{>{\hfil$}p{3cm}<{$\hfil}}
\newcolumntype{P}{>{\hfil$}p{7.7cm}<{$\hfil}}
\newcolumntype{F}{>{\hfil$}p{5.7cm}<{$\hfil}}
\newcolumntype{S}{>{\hfil$}p{1.8cm}<{$\hfil}}
\newcolumntype{R}{>{\hfil$}p{5.2cm}<{$\hfil}}
\newcolumntype{U}{>{\hfil$}p{4.2cm}<{$\hfil}}
\newcolumntype{Q}{>{\hfil$}p{6.4cm}<{$\hfil}}
\newcolumntype{T}{>{\hfil$}p{1.9cm}<{$\hfil}}
\newcolumntype{V}{>{\hfil$}p{5.8cm}<{$\hfil}}
\newcolumntype{H}{>{\hfil$}p{1.8cm}<{$\hfil}}
\newcolumntype{A}{>{\hfil$}p{6cm}<{$\hfil}}
\newcolumntype{B}{>{\hfil$}p{2cm}<{$\hfil}}

\makeatletter
\newcommand\xleftrightarrow[2][]{%
  \ext@arrow 9999{\longleftrightarrowfill@}{#1}{#2}}
\newcommand\longleftrightarrowfill@{%
  \arrowfill@\leftarrow\relbar\rightarrow}
\makeatother

\numberwithin{equation}{section}

\definecolor{cambridgeblue}{rgb}{0.64, 0.76, 0.68}
\definecolor{caribbeangreen}{rgb}{0.0, 0.8, 0.6}
\definecolor{celadon}{rgb}{0.67, 0.88, 0.69}
\definecolor{champagne}{rgb}{0.97, 0.91, 0.81}
\definecolor{cream}{rgb}{1.0, 0.99, 0.82}
\definecolor{cyan(process)}{rgb}{0.0, 0.72, 0.92}
\definecolor{brilliantlavender}{rgb}{0.96, 0.73, 1.0}
\definecolor{candypink}{rgb}{0.89, 0.44, 0.48}

\usepackage{pdflscape}

\def\vQ#1#2{\overset{|}{\mathsf{Q}}_{#1,#2}}
\def\hQ#1#2{\overset{-}{\mathsf{Q}}_{#1,#2}}
\def\bQ#1#2{\overset{\backslash}{\mathsf{Q}}_{#1,#2}}
\def\sQ#1#2{\overset{\slash}{\mathsf{Q}}_{#1,#2}}

\usetikzlibrary{positioning}
\usetikzlibrary{arrows}
\usetikzlibrary{decorations.pathreplacing}
\usetikzlibrary{shapes.geometric}
\usetikzlibrary{calc}
\usetikzlibrary{decorations.pathmorphing}
\usetikzlibrary{shapes.misc}

\tikzset{gaugeSU/.style={inner sep=1.7mm,draw=none,fill=yellow,minimum size=2mm,circle, draw}}
\tikzset{flavourSU/.style={draw=none,minimum size=2mm,fill=white, regular polygon,regular polygon sides=4,draw}}
\tikzset{flavour/.style={draw=none,minimum size=0.3mm,fill=white, regular polygon,regular polygon sides=4,draw}}
\tikzset{gaugeBig/.style={inner sep=1.7mm,draw=none,fill=white,minimum size=2mm,circle, draw}}
\tikzset{bd/.style={circle, draw=black, inner sep=0pt, fill=black, minimum size=2mm}}
\tikzset{wd/.style={circle, draw=black, inner sep=0pt, fill=white, minimum size=2mm}}
\tikzset{Dynkin/.style={circle, draw=black, inner sep=0pt, fill=white, minimum size=2mm}}
\tikzstyle{ligne}=[draw, very thick] 
\tikzstyle{gridline}=[draw, gray] 
\tikzset{gauge/.style={circle, draw,inner sep=2.5pt}}
\tikzset{gaugeo/.style={circle, draw,inner sep=2.5pt,fill=orange}}
\tikzset{gaugec/.style={circle, draw,inner sep=2.5pt,fill=cyan}}
\tikzset{gauger/.style={circle, draw,inner sep=2.5pt,fill=red}}
\tikzset{gaugeb/.style={circle, draw,inner sep=2.5pt,fill=blue}}
\tikzset{gaugeg/.style={circle, draw,inner sep=2.5pt,fill=green}}
\tikzset{gaugem/.style={circle, draw,inner sep=2.5pt,fill=magenta}}
\tikzset{gaugey/.style={circle, draw,inner sep=2.5pt,fill=yellow}}
\tikzset{hasse/.style={circle, fill,inner sep=2pt}}
\tikzset{shrinky/.style={circle, fill,inner sep=1pt}}
\tikzset{sized/.style={circle, draw, inner sep=1.5pt}}
\tikzset{seven/.style={circle, draw,inner sep=3pt}}

\tikzset{dotto/.style={circle, orange, draw,inner sep=1.5pt,fill=orange}}
\tikzset{dottp/.style={circle, purple, draw,inner sep=1.5pt,fill=purple}}
\tikzset{dottc/.style={circle, cyan, draw,inner sep=1.5pt,fill=cyan}}
\tikzset{dottr/.style={circle, red, draw,inner sep=1.5pt,fill=red}}
\tikzset{dottb/.style={circle, blue, draw,inner sep=1.5pt,fill=blue}}
\tikzset{dottg/.style={circle, green, draw,inner sep=1.5pt,fill=green}}
\tikzset{dottm/.style={circle, magenta, draw,inner sep=1.5pt,fill=magenta}}

\usepackage{ifthen}
\usepackage{xstring}

\usepackage{cancel}
\newcommand{\xdownarrow}[1]{%
  {\left\downarrow\vbox to #1{}\right.\kern-\nulldelimiterspace}
}

\begin{document}

\begin{titlepage}

\phantom{wowiezowie}

\vspace{-1cm}

\begin{center}

{\Huge {\bf 5d Higgs Branches:}}

\bigskip

{\Huge {\bf Stratifications from Geometry}}

\vspace{1cm}

{\Large  Mario De Marco,$^{\sharp}$ Michele Del Zotto,$^{\dagger\ddagger\star}$}\\ 

\medskip

{\Large  Julius F. Grimminger,$^{\flat}$ and Andrea Sangiovanni $^{\dagger\ddagger\star}$}\\

\vspace{1cm}

{\it
{\small

$^\dagger$ Mathematics Institute, Uppsala University, \\ Box 480, SE-75106 Uppsala, Sweden\\
\vspace{.25cm}
$^\star$ Centre for Geometry and Physics, Uppsala University, \\ Box 480, SE-75106 Uppsala, Sweden\\
\vspace{.25cm}
$^\ddagger$ Department of Physics and Astronomy, Uppsala University,\\ Box 516, SE-75120 Uppsala, Sweden\\
\vspace{.25cm}
$^{\flat}$ Mathematical Institute, University of Oxford,\\Andrew Wiles Building, Woodstock Road, Oxford, OX2 6GG, UK\\
\vspace{.25cm}
$^{\sharp}$  Physique Th\'eorique et Math\'ematique and International Solvay Institutes\\
Universit\'e Libre de Bruxelles, C.P. 231, 1050 Brussels, Belgium
}}

\vskip .5cm
{\footnotesize \tt mario.de.marco@ulb.be \hspace{1cm} michele.delzotto@math.uu.se } \\
{\footnotesize \tt    julius.grimminger@maths.ox.ac.uk \hspace{1cm} andrea.sangiovanni@math.uu.se}

\vskip 1cm
     	{\bf Abstract }
\vskip .1in

\end{center}

\noindent Higgs Branches of 5d SCFTs are hyperkähler cones, symplectic varieties with interesting nested singularities governing the structure of possible Higgs branch RG flows. The main purpose of this note is to investigate 5d Higgs Branches via a geometric route: we study 5d SCFTs geometrically engineered from M-theory on non-compact Calabi-Yau threefolds, with non-isolated singularities. We match the (dynamical) complex structure deformations of the threefold with the symplectic leaves in the Higgs branch of the 5d SCFT, constructing its stratification purely relying on geometry. Consistency checks for our proposal are presented exploiting magnetic quivers.

\eject

\end{titlepage}

\tableofcontents

\section{Introduction}

In the mid-nineties it was realized that superconformal interacting field theories (SCFTs) exist in dimension greater than four \cite{Witten:1995ex,Strominger:1995ac,Witten:1995em,Ganor:1996mu,Seiberg:1996qx,Seiberg1996,Morrison_1997,Douglas:1996xp}. All of these higher-dimensional fixed points do not have conventional Lagrangian descriptions, which required developing unconventional tools to capture their properties. This is one of the main reason to develop \textit{geometric engineering} techniques within string theory -- a perspective advocated in the seminal work by Leung and Vafa \cite{leung1997branes}. Geometric engineering is a top-down technology that employs the dimensional reduction of M-/F-/string theory on a non-compact singular variety, thus decoupling gravitational degrees of freedom and realizing a field theory in the transverse directions. This method evades traditional constraints imposed by perturbative field theory, since it can extract salient data directly at the strongly-coupled conformal fixed point, with no need of a corresponding weakly coupled Lagrangian description. Our main focus in this work is to develop geometric engineering tools to capture the structure of symplectic singularities along the Higgs branch of 5d SCFTs geometrically engineered in M-theory. In particular we introduce the geometric Hasse diagram as an alternative method to recover the stratification characterizing Higgs branch RG flows.

\medskip

It is well-known that 5d SCFTs can be engineered by M-theory on a non-compact Calabi-Yau threefold with canonical singularities \cite{Witten_1996,Douglas:1996xp,Morrison_1997,Intriligator_1997} (see also e.g. \cite{Xie_2017,Jefferson_2018,Bhardwaj:2019jtr,Bhardwaj:2019ngx,Closset_2019,Apruzzi:2019opn,Apruzzi_2020}). Among the key features that can be probed via geometric engineering a prominent role is played by the moduli spaces of vacua of the 5d SCFTs. Due to supersymmetry, the moduli space of these models admits Coulomb, Higgs and mixed branches. A firm understanding of the physics along the Coulomb Branch (CB) is by now established \cite{Intriligator_1997,Morrison_1997,Jefferson:2017ahm}. The Coulomb branch (CB) is a real cone of dimension $r$, with $r$ the rank of the theory, and that the low-energy dynamics on a generic point of the CB is captured by a prepotential \cite{Seiberg:1996bd}. It is often useful to consider the extended CB of the model, which is a fibration of the CB over the susy mass deformations of the theory -- for a discussion see e.g.\cite{Xie_2017,Closset_2019} -- where gauge theory phases are realized. In the context of 5d SCFTs engineered by M-theory on a non-compact Calabi-Yau singular threefold, the extended CB is realized via the extended K\"ahler cone of the singularity \cite{Witten:1996qb}. The coefficient of the 5d CB prepotential are captured by the triple intersection number between the compact divisors, for a given resolution of the singularity. Different chambers along the CB corresponds to birational equivalent resolutions (massless particles giving rise to the transitions are realized by M2 branes wrapping vanishing BPS 2-cycles) and boundaries of the CB correspond to loci where divisors shrink to zero size (tensionless BPS strings). The resolution of a canonical singularity is, despite often being technically subtle, well-understood from the mathematical point of view. This guarantees full control on the physics along the Coulomb Branch. In particular, along the extended Coulomb branch of a 5d SCFT various gauge theory phases can emerge, with Higgs branches of their own (often dubbed Higgs branches at finite coupling in the literature).\\
\indent A completely different story pertains to the Higgs branch of the SCFT: it is a hyperk\"ahler cone, which naturally comes stratified in symplectic leaves. Physically, each leaf corresponds to a different set of massless excitations, which often do not have a Lagrangian description, leading to interacting SCFTs. From the geometric point of view, the Higgs Branch is described by the dynamical complex structure deformations of the Calabi-Yau threefold employed to engineer the 5d SCFT, corresponding to the spectrum of the 5d Higgs branch chiral ring. Ideally, one would like to characterize the Higgs Branch in terms of the following data (in increasing degree of technical difficulty):
\begin{itemize}
    \item its quaternionic dimension at the conformal point;
    \item its stratification in terms of symplectic leaves, computing the dimension of each leaf;
    \item its chiral ring;
    \item its metric.
\end{itemize}
Unluckily, the deformation theory of Calabi-Yau threefolds is known only in a handful of cases, such as isolated singularities, where the Milnor ring can be straightforwardly computed. This gains access to the Higgs Branch dimension. Even more direly, the deformation theory of threefolds with non-isolated singularities has only been partially explored so far (see e.g.\ \cite{siersma2001vanishing} and related references).\\
\indent What can be said for the other items in the list? Up to this note, the gold standard for the characterization of the Higgs Branch has been the \textit{magnetic quiver (MQ) technology}, which captures the Higgs Branch of some 5d SCFT in terms of the Coulomb Branch of a 3d $\mathcal{N}=4$ MQ theory \cite{Cabrera:2018jxt}.\footnote{More broadly, the MQ is well-defined for all theories with 8 supercharges in dimension greater or equal than three.} The stratification can be efficiently outlined via quiver subtraction or the decay and fission algorithm \cite{Cabrera:2018ann,Cabrera:2018jxt,Bourget:2023dkj,Bourget:2024mgn}. The Higgs branch chiral ring corresponds to the ring of dressed monopole operators in 3d, which can be computed through the monopole formula \cite{Cremonesi:2013lqa}. The magnetic quiver can be algorithmically written down when the CY3 employed to engineer the 5d SCFT corresponds to a 5-brane web\footnote{E.g.\ toric threefolds, or threefolds corresponding to generalized toric polygons (GTPs). The MQ is also known for brane-webs with orientifold planes, which have no known explicit dual CY3.}.\\

\indent In this work, \textit{we study the stratification of Higgs Branches of 5d SCFTs from a purely geometric point of view, employing methods rooted in M-theory geometric engineering}. We consider classes of toric CY3 with non-isolated singularities, and we explicitly compute their dynamical deformations\footnote{Recently, it has been argued on mathematical grounds that very simple geometries such as the conifold do not admit any normalizable 3-form. We believe this is related to the fact that the uniqueness of the CY metrics is no longer true for non-compact CY 3-folds. In particular, this could entail that singularities with the same algebraic structure admit different CY metrics, for which the dynamical deformations are different, which would explain the observation in \cite{Acharya:2024bnt}. In this work we momentarily pause this discussion, and we focus primarily on showing agreement between the MQ and the geometric perspective, making use of an effective definition of ``dynamical'' deformations. Once this correspondence is firmly established in general, the question of which deformations can be rigorously defined as normalizable will come into the foreground.}.
We compute the UV Higgs Branch dimension for the corresponding SCFTs, and inspect its stratification, reconstructing their full Hasse diagram: the chief technical resource at our disposal are crepant resolutions of the engineering threefolds, and a thorough understanding of the complex deformations at the conformal point. We show complete agreement with the results on the stratification of the Higgs Branch expected from magnetic quiver techniques. Our work therefore presents evidence that geometric methods can be exploited towards characterizing fully not only the HB dimension but also its stratification in symplectic singularities, goverining HB RG flows. Further work is necessary to extract from geometry the chiral ring and HB metric. Nevertheless, our result is suggestive of an identification of monomial deformations on the geometric side with monopole vevs in the magnetic quiver, which is discussed in our forthcoming publication \cite{toappear}.

\medskip

A natural question arising from our work pertains the breadth of its application in the realm of toric threefolds, in view of the study of the Higgs branch of the related 5d SCFTs. In general, toric threefolds can possess both isolated and non-isolated singularities: for the isolated ones, the algorithm by Altmann \cite{altmann2000one,altmann1994versal} provides an efficient way to explicitly compute the complex structure deformations. We expect that Altmann's method, suitably combined with the proposal presented here, would allow to tackle the derivation of dynamical complex structure deformations for an arbitrary toric threefold in a purely geometric fashion. Furthermore, it would be interesting to extend our analysis outside the toric realm, addressing the cases in which a brane-web is still available to describe the 5d SCFT (see e.g.\ the examples analyzed in \cite{DeMarco:2025ugw}).\\  

We do not delve into the role of M2-brane instanton corrections to the Higgs Branch, as these only affect its metric, and leave its UV dimension and stratification unscathed. 

\subsection*{Structure of the work}

\indent We structure the work as follows: in Section \ref{sec:stratificationfromgeometry} we lay out our philosophy, introducing geometric Hasse diagrams, which encode the stratification of 5d Higgs branches from a purely geometric point of view. We also show our method to capture key features of the Higgs Branch in terms of explicit dynamical complex structure deformations of the threefold. In Section \ref{sec: box diagrams} we show our main working examples, which are a class of toric CY3 with non-isolated singularities, whose toric diagram is a rectangle. We compute the stratitification of its Higgs branch from geometry, and check its full compatibility with the magnetic quiver results giving a detailed thorough analysis for the simplest non-trivial model in this class, the $\mathcal T_{3,3}$ 5d SCFT. In Section \ref{sec: Tn theories} we apply our results to the 5d $T_n$ theories, discussing the details of the $\mathcal T_4$ theory as an example. Finally, in Section \ref{sec:white dots} we connect our results to the literature on GTPs, that comprise a subset of the full list of dynamical deformations. In Appendix \ref{sec:complete intersections} we sketch the generalization of our method to broader classes of toric CY3.

\section{Higgs branch stratification from geometry}
\label{sec:stratificationfromgeometry}

\subsection{The Geometric Hasse diagram}
We operate in the context of M-theory geometric engineering on a non-compact canonical threefold, which is thought to produce a 5d superconformal theory in the transverse directions \cite{Xie_2017}. Our main proposal is as follows: we define a new tool to investigate Higgs branches of 5d SCFTs, purely in geometrical terms,the \textit{Geometric Hasse diagram}. The geometric Hasse diagram is a property of the threefold $X$ corresponding to a  5d SCFT $\mathcal{T}$ via geometric engineering which encodes the stratification of the Higgs branch in terms of the geometry of a subset of the corresponding complex structure deformations, the \textit{dynamical deformations}. The geometric Hasse diagram is particularly nice for those models which are engineered by threefold with non-isolated singularities, such as the ones arising in the context of the atomic classification \cite{DeMarco:2023irn,DeMarco:2025ugw,Bourget:2026ono,DeMarco:2026tnc}. In this case, the non-isolated singularities can be exploited to organize and identify the dynamical deformations among the complex structure deformations. This gives a coarse hierarchy which can be exploited as a guide towards the full stratification of the Higgs branch of the SCFT. 

\medskip

For the sake of concreteness and simplicity in this work, we focus on 5d SCFTs defined by a toric threefold with non-isolated singularities which is either a hypersurface in $\mathbb{C}^4$ spanned by 
\begin{equation}\label{eq:coord1}
\mathbb C^4 = \text{Spec }\mathbb C[x,y,z,w]
\end{equation}
or a complete intersection of two equations in $\mathbb{C}^5$ spanned by 
\begin{equation}\label{eq:coord2}
\mathbb C^5 = \text{Spec }\mathbb C[x,y,z,u,v].
\end{equation}
Relying on the standard geometric engineering dictionary, we know all the details of the crepant resolution of such threefolds, which captures the Coulomb Branch as well as the flavor symmetry of $\mathcal{T}$. For such geometries, we characterize \textit{all} the dynamical deformations of the theory, which parametrize its Higgs branch, building on the technology introduced in \cite{Collinucci:2020jqd}. More precisely, we show that the dynamical deformations that we characterize encompass the full Higgs branch dimension and contain the ones found in \cite{Collinucci:2020jqd} and \cite{Alexeev:2024bko}. In particular, we identify additional dynamical deformations that are not comprised in the frameworks cited above and are crucial to recover the SCFT Higgs branch.

\medskip

The data encoded by the geometric Hasse diagram of $X$ is as follows:
\begin{itemize}
    \item Each leaf of the Hasse diagram is encoded by an explicit canonical threefold (not necessarily toric, as a result of the dynamical deformations). Such singularity corresponds via geometric engineering to a specific 5d SCFT, and as such it defines an equivalence class --- in general, a given 5d SCFT can be equivalently described by different threefolds (we spell out such equivalence criterion for threefolds momentarily). All the data that can be extracted from the geometric engineering dictionary is then fully accessible: even if the singularities are generically non-toric, both the crepant resolution and the unobstructed dynamical deformations can be computed. Leaves correspond to nodes of the geometric Hasse diagrams.
    \item Dynamical complex structure deformations of the engineering threefolds appear as edges in the Hasse diagram, and correspond via geometric engineering to Higgs branch RG flows.
\end{itemize}
In the toric threefolds at hand, the dynamical deformations are generally grouped into families of monomials whose members depend on the specific presentation of the singularity. More specifically, in what follows, we will concentrate on two classes of examples: 
\begin{enumerate}
\item The first class is 
\begin{equation}
    X_{n,k}=\left\{(x,y,u,v,z)\in\mathbb{C}^5\left|\begin{matrix}
        xy=z^n\\
        uv=z^k
    \end{matrix}\right.\right\}\;.
\end{equation} and the corresponding dynamical deformations are monomials out of the following families of variables
\begin{equation}
\label{eq:familiesrect}
(x,z), \quad (y,z), \quad (u,z), \quad (v,z), \quad (z)\,.
\end{equation}

\item The second class is 
\begin{equation}\label{Tn hypersurface sec 2}
    xyz = w^n \quad \subset \mathbb{C}^4
\end{equation}
and the corresponding dynamical deformations are monomials out of the following families of variables
\begin{equation}
\label{eq:familieshypersurf}
     (x,w), \quad (y,w), \quad (z,w), \quad (w),
\end{equation}

\end{enumerate}
This neat organization of dynamical deformations hinges on the presence of a symmetric pattern of non-isolated singularities in the threefold $X$ that engineers $\mathcal{T}$.

\medskip

Given a choice of $\mathcal{T}$ engineered by M-theory on $X$ with the properties above, its \textit{geometric Hasse diagram} is realized as follows:
\begin{itemize}
    \item The top leaf of the Hasse diagram is denoted by the theory $\mathcal{T}$ (corresponding to the singularity $X$);
    \item Every other leaf can be reached by turning on a (combination of) dynamical deformations of the threefold $X$. Hence, edges in the Hasse diagram are encoded by the dynamical deformations of the original threefold.
    \item The singularities corresponding to \textit{leaves} of the Hasse diagram $X_i$ are characterized by the following invariants:
    \begin{itemize}
        \item $\dim H_4(\hat X_i)$ (where $\hat X_i$ is a crepant resolution of $X_i$), i.e. the rank of the Coulomb branch of the theory $\mathcal T_{X_i}$
        \item $\dim H_2(\hat X_i)-\dim H_4(\hat X_i)$, i.e. the rank of the flavor symmetry of the theory $\mathcal T_{X_i}$
        \item the number of independent dynamical deformations, i.e. the dimension of the Higgs branch of the corresponding SCFT
    \end{itemize}
    This data is not sufficient to univocally encode a leaf: theories engineered by distinct threefolds (say $X_a$ and $X_b$), for which the data above is identical, can still lie on distinct leaves. What distinguishes them is the following geometric principle: 
    
    \textit{two 5d SCFTs engineered by different threefolds $X_a$ and $X_b$, with identical CB and flavor symmetry rank (encoded by their resolution), and identical Higgs branch (encoded by their dynamical deformations) correspond to the same leaf if and only if they are engineered by the initial theory singularity through dynamical deformations that belong to the same combination of families  of the dynamical deformations in \eqref{eq:familiesrect} and \eqref{eq:familieshypersurf}}.\footnote{This requirement is needed in order to distinguish symmetric branches of the Hasse diagram. Explicitly, if $X_a$ is defined by deformations o in the family $S_a$ and $X_b$ is encoded by $S_b$, then we require that $S_a \subseteq S_b$, upon possible relabeling of the threefolds.}
    
    The rationale underlying this principle is that two threefolds $X_a$ and $X_b$ can engineer 5d SCFTs on the same leaf only if the deformations that modify the parent threefold $X$ into $X_a$ and $X$ into $X_b$ break the \textit{same non-isolated singular lines}, which are precisely encoded by the deformations in \eqref{eq:familiesrect} and \eqref{eq:familieshypersurf}.
\end{itemize}
In this fashion, we are able to check in concrete examples that the geometric Hasse diagram of $\mathcal{T}$ precisely matches the Higgs branch Hasse diagram derived from the magnetic quiver technology \cite{Cabrera:2018ann,Cabrera:2018jxt,Bourget:2023dkj,Bourget:2024mgn}. Furthermore, we match each dynamical complex structure deformation of the threefold to its avatar in term of movements of 5-brane subwebs in the corresponding brane description of the same 5d SCFT. We expect that a suitable generalization of the definitions presented in this Section holds also for more general threefold singularities. We leave this challenge for future work.

\section{The $\mathcal{T}_{3,3}$ theory: a detailed analysis}\label{sec: box diagrams}

In this Section we discuss an example of a particularly symmetric model, the $\mathcal{T}_{3,3}$ theory where the whole stratification Hasse diagram can be reconstructed out of the mixed Higgs branches along the corresponding gauge theory phases.

\subsection{The $\mathcal T_{n,k}$ models and their gauge theory phases}

The $\mathcal{T}_{3,3}$ theory belongs to a nice family of 5d $\mathcal{N}=1$ SCFTs geometrically engineered in M-theory on $\mathbb{R}^{(1,4)}\times X_{n,k}$, where $X_{n,k}$ is a toric local CY3 singularity encoded by the following toric diagram \cite{leung1997branes}
\begin{equation}\label{eq:boxdiagram}
\begin{gathered}
    \begin{tikzpicture}
        \node[bd] (00) at (0,0) {};
        \node[bd] (10) at (1,0) {};
        \node (20) at (2,0) {$\cdots$};
        \node[bd] (30) at (3,0) {};
        \node[bd] (40) at (4,0) {};
        \node[bd] (04) at (0,4) {};
        \node[bd] (14) at (1,4) {};
        \node (24) at (2,4) {$\cdots$};
        \node[bd] (34) at (3,4) {};
        \node[bd] (44) at (4,4) {};
        \node[bd] (01) at (0,1) {};
        \node (02) at (0,2) {$\vdots$};
        \node[bd] (03) at (0,3) {};
        \node[bd] (41) at (4,1) {};
        \node (42) at (4,2) {$\vdots$};
        \node[bd] (43) at (4,3) {};
        \draw (00)--(10)--(20)--(30)--(40)--(41)--(42)--(43)--(44)--(34)--(24)--(14)--(04)--(03)--(02)--(01)--(00);
        \draw [decorate,decoration={brace,amplitude=5pt},xshift=0pt,yshift=0pt] (4.1,-0.3)--(-0.1,-0.3) node [black,midway,yshift=-0.5cm] {$n$ edges};
        \draw [decorate,decoration={brace,amplitude=5pt},xshift=0pt,yshift=0pt] (-0.3,-0.1)--(-0.3,4.1) node [black,midway,xshift=-0.5cm,rotate=90] {$k$ edges};
    \end{tikzpicture}\;.
    \end{gathered}
\end{equation}
We label the corresponding 5d SCFTs, $\mathcal T_{n,k}$ theories. $X_{n,k}$ is a complete intersection in $\mathbb{C}^5$ defined by two equations,
\begin{equation}\label{fig: box diagram}
    X_{n,k}=\left\{(x,y,u,v,z)\in\mathbb{C}^5\left|\begin{matrix}
        xy=z^n\\
        uv=z^k
    \end{matrix}\right.\right\}\;.
\end{equation}
Clearly $\mathcal{T}_{n,k}=\mathcal{T}_{k,n}$ since $X_{n,k}=X_{k,n}$. Without loss of generality in the following we require $n\leq k$.

\medskip

There are four distinct gauge theory phases for this singularity, obtained by mass deformations of $\mathcal{T}_{n,k}$ and realized alone the extended Coulomb branch of the model. These are encoded by four distinct rulings \cite{Closset_2019} of $X_{n,k}$ (see also \cite{Bhardwaj:2019ngx}). For our purposes we only need to focus on the following two:
\begin{equation}
    \vQ{n}{k}=\quad\raisebox{-.8\height}{\scalebox{0.7}{\begin{tikzpicture}
        \node[flavourSU,label=below:{$k$}] (0) at (0,0) {};
        \node[gaugeSU,label=below:{$k$}] (1) at (1,0) {};
        \node (2) at (2,0) {$\cdots$};
        \node[gaugeSU,label=below:{$k$}] (3) at (3,0) {};
        \node[flavourSU,label=below:{$k$}] (4) at (4,0) {};
        \draw (0)--(1)--(2)--(3)--(4);
        \draw [decorate,decoration={brace,amplitude=5pt},xshift=0pt,yshift=0pt] (3.1,-0.8)--(1-0.1,-0.8) node [black,midway,yshift=-0.5cm] {$n-1$ nodes};
    \end{tikzpicture}}}\qquad\textnormal{realised on}\qquad\raisebox{-.6\height}{\scalebox{0.7}{\begin{tikzpicture}
        \node[bd] (00) at (0,0) {};
        \node[bd] (10) at (1,0) {};
        \node (20) at (2,0) {$\cdots$};
        \node[bd] (30) at (3,0) {};
        \node[bd] (40) at (4,0) {};
        \node[bd] (04) at (0,4) {};
        \node[bd] (14) at (1,4) {};
        \node (24) at (2,4) {$\cdots$};
        \node[bd] (34) at (3,4) {};
        \node[bd] (44) at (4,4) {};
        \node[bd] (01) at (0,1) {};
        \node (02) at (0,2) {$\vdots$};
        \node[bd] (03) at (0,3) {};
        \node[bd] (41) at (4,1) {};
        \node (42) at (4,2) {$\vdots$};
        \node[bd] (43) at (4,3) {};
        \draw (00)--(10)--(20)--(30)--(40)--(41)--(42)--(43)--(44)--(34)--(24)--(14)--(04)--(03)--(02)--(01)--(00);
        \draw [decorate,decoration={brace,amplitude=5pt},xshift=0pt,yshift=0pt] (4.1,-0.3)--(-0.1,-0.3) node [black,midway,yshift=-0.5cm] {$n$ edges};
        \draw [decorate,decoration={brace,amplitude=5pt},xshift=0pt,yshift=0pt] (-0.3,-0.1)--(-0.3,4.1) node [black,midway,xshift=-0.5cm,rotate=90] {$k$ edges};
        \node[bd] (11) at (1,1) {};
        \node (12) at (1,2) {$\vdots$};
        \node[bd] (13) at (1,3) {};
        \node[bd] (31) at (3,1) {};
        \node (32) at (3,2) {$\vdots$};
        \node[bd] (33) at (3,3) {};
        \draw (10)--(11)--(12)--(13)--(14) (30)--(31)--(32)--(33)--(34);
    \end{tikzpicture}}}\;,
\end{equation}
and
\begin{equation}
    \hQ{n}{k}=\quad\raisebox{-.5\height}{\scalebox{0.7}{\begin{tikzpicture}
        \node[flavourSU,label=left:{$n$}] (0) at (0,0) {};
        \node[gaugeSU,label=left:{$n$}] (1) at (0,1) {};
        \node (2) at (0,2) {$\vdots$};
        \node[gaugeSU,label=left:{$n$}] (3) at (0,3) {};
        \node[flavourSU,label=left:{$n$}] (4) at (0,4) {};
        \draw (0)--(1)--(2)--(3)--(4);
        \draw [decorate,decoration={brace,amplitude=5pt},xshift=0pt,yshift=0pt] (-0.8,1-0.1)--(-0.8,3.1) node [black,midway,xshift=-0.5cm,rotate=90] {$k-1$ nodes};
    \end{tikzpicture}}}\qquad\textnormal{realised on}\qquad\raisebox{-.6\height}{\scalebox{0.7}{\begin{tikzpicture}
        \node[bd] (00) at (0,0) {};
        \node[bd] (10) at (1,0) {};
        \node (20) at (2,0) {$\cdots$};
        \node[bd] (30) at (3,0) {};
        \node[bd] (40) at (4,0) {};
        \node[bd] (04) at (0,4) {};
        \node[bd] (14) at (1,4) {};
        \node (24) at (2,4) {$\cdots$};
        \node[bd] (34) at (3,4) {};
        \node[bd] (44) at (4,4) {};
        \node[bd] (01) at (0,1) {};
        \node (02) at (0,2) {$\vdots$};
        \node[bd] (03) at (0,3) {};
        \node[bd] (41) at (4,1) {};
        \node (42) at (4,2) {$\vdots$};
        \node[bd] (43) at (4,3) {};
        \draw (00)--(10)--(20)--(30)--(40)--(41)--(42)--(43)--(44)--(34)--(24)--(14)--(04)--(03)--(02)--(01)--(00);
        \draw [decorate,decoration={brace,amplitude=5pt},xshift=0pt,yshift=0pt] (4.1,-0.3)--(-0.1,-0.3) node [black,midway,yshift=-0.5cm] {$n$ edges};
        \draw [decorate,decoration={brace,amplitude=5pt},xshift=0pt,yshift=0pt] (-0.3,-0.1)--(-0.3,4.1) node [black,midway,xshift=-0.5cm,rotate=90] {$k$ edges};
        \node[bd] (11) at (1,1) {};
        \node (21) at (2,1) {$\cdots$};
        \node[bd] (31) at (3,1) {};
        \node[bd] (13) at (1,3) {};
        \node (23) at (2,3) {$\cdots$};
        \node[bd] (33) at (3,3) {};
        \draw (01)--(11)--(21)--(31)--(41) (03)--(13)--(23)--(33)--(43);
    \end{tikzpicture}}}\;.
\end{equation}
The other correspond to (we assume $n\leq k$):
\begin{equation}
\begin{split}
    \bQ{n}{k}=\quad\raisebox{-.8\height}{\scalebox{0.7}{\begin{tikzpicture}
        \node[gaugeSU,label=below:{$1$}] (0) at (0,0) {};
        \node[gaugeSU,label=below:{$2$}] (1) at (1,0) {};
        \node (2) at (2,0) {$\cdots$};
        \node[gaugeSU,label=below:{$n-1$}] (3) at (3,0) {};
        \node[gaugeSU,label=below:{$(n)_{-\frac{1}{2}}$}] (4) at (4.3,0) {};
        \node[gaugeSU,label=below:{$n$}] (5) at (5.3,0) {};
        \node (6) at (6.3,0) {$\cdots$};
        \node[gaugeSU,label=below:{$n$}] (7) at (7.3,0) {};
        \node[gaugeSU,label=below:{$(n)_{\frac{1}{2}}$}] (8) at (8.3,0) {};
        \node[gaugeSU,label=below:{$n-1$}] (9) at (9.6,0) {};
        \node (10) at (10.6,0) {$\cdots$};
        \node[gaugeSU,label=below:{$2$}] (11) at (11.6,0) {};
        \node[gaugeSU,label=below:{$1$}] (12) at (12.6,0) {};
        \draw (0)--(1)--(2)--(3)--(4)--(5)--(6)--(7)--(8)--(9)--(10)--(11)--(12);
        \draw [decorate,decoration={brace,amplitude=5pt},xshift=0pt,yshift=0pt] (12.6+0.1,-1)--(0-0.1,-1) node [black,midway,yshift=-0.5cm] {$n+k-1$ nodes};
    \end{tikzpicture}}}\\
    \textnormal{realised on}\qquad\raisebox{-.6\height}{\scalebox{0.7}{\begin{tikzpicture}
        \node[bd] (00) at (0,0) {};
        \node[bd] (10) at (1,0) {};
        \node (20) at (2,0) {$\cdots$};
        \node[bd] (30) at (3,0) {};
        \node[bd] (40) at (4,0) {};
        \node[bd] (04) at (0,4) {};
        \node[bd] (14) at (1,4) {};
        \node (24) at (2,4) {$\cdots$};
        \node[bd] (34) at (3,4) {};
        \node[bd] (44) at (4,4) {};
        \node[bd] (01) at (0,1) {};
        \node (02) at (0,2) {$\vdots$};
        \node[bd] (03) at (0,3) {};
        \node[bd] (41) at (4,1) {};
        \node (42) at (4,2) {$\vdots$};
        \node[bd] (43) at (4,3) {};
        \draw (00)--(10)--(20)--(30)--(40)--(41)--(42)--(43)--(44)--(34)--(24)--(14)--(04)--(03)--(02)--(01)--(00);
        \draw [decorate,decoration={brace,amplitude=5pt},xshift=0pt,yshift=0pt] (4.1,-0.3)--(-0.1,-0.3) node [black,midway,yshift=-0.5cm] {$n$ edges};
        \draw [decorate,decoration={brace,amplitude=5pt},xshift=0pt,yshift=0pt] (-0.3,-0.1)--(-0.3,4.1) node [black,midway,xshift=-0.5cm,rotate=90] {$k$ edges};
        \node[bd] (11) at (1,1) {};
        \node[bd] (13) at (1,3) {};
        \node[bd] (31) at (3,1) {};
        \node[bd] (33) at (3,3) {};
        \def\x{0.3};
        \draw (01)--(10) (03)--(0+\x,3-\x) (04)--(13)--(1+\x,3-\x) (1-\x,1+\x)--(11)--(1+\x,1-\x) (14)--(1+\x,4-\x) (3-\x,0+\x)--(30) (3-\x,1+\x)--(31)--(40) (3-\x,3+\x)--(33)--(3+\x,3-\x) (34)--(43) (4-\x,1+\x)--(41);
    \end{tikzpicture}}}\;,
\end{split}
\end{equation}
and
\begin{equation}
\begin{split}
    \sQ{n}{k}=\quad\raisebox{-.8\height}{\scalebox{0.7}{\begin{tikzpicture}
        \node[gaugeSU,label=below:{$1$}] (0) at (0,0) {};
        \node[gaugeSU,label=below:{$2$}] (1) at (1,0) {};
        \node (2) at (2,0) {$\cdots$};
        \node[gaugeSU,label=below:{$n-1$}] (3) at (3,0) {};
        \node[gaugeSU,label=below:{$(n)_{-\frac{1}{2}}$}] (4) at (4.3,0) {};
        \node[gaugeSU,label=below:{$n$}] (5) at (5.3,0) {};
        \node (6) at (6.3,0) {$\cdots$};
        \node[gaugeSU,label=below:{$n$}] (7) at (7.3,0) {};
        \node[gaugeSU,label=below:{$(n)_{\frac{1}{2}}$}] (8) at (8.3,0) {};
        \node[gaugeSU,label=below:{$n-1$}] (9) at (9.6,0) {};
        \node (10) at (10.6,0) {$\cdots$};
        \node[gaugeSU,label=below:{$2$}] (11) at (11.6,0) {};
        \node[gaugeSU,label=below:{$1$}] (12) at (12.6,0) {};
        \draw (0)--(1)--(2)--(3)--(4)--(5)--(6)--(7)--(8)--(9)--(10)--(11)--(12);
        \draw [decorate,decoration={brace,amplitude=5pt},xshift=0pt,yshift=0pt] (12.6+0.1,-1)--(0-0.1,-1) node [black,midway,yshift=-0.5cm] {$n+k-1$ nodes};
    \end{tikzpicture}}}\\
    \textnormal{realised on}\qquad\raisebox{-.6\height}{\scalebox{0.7}{\begin{tikzpicture}
        \node[bd] (00) at (0,0) {};
        \node[bd] (10) at (1,0) {};
        \node (20) at (2,0) {$\cdots$};
        \node[bd] (30) at (3,0) {};
        \node[bd] (40) at (4,0) {};
        \node[bd] (04) at (0,4) {};
        \node[bd] (14) at (1,4) {};
        \node (24) at (2,4) {$\cdots$};
        \node[bd] (34) at (3,4) {};
        \node[bd] (44) at (4,4) {};
        \node[bd] (01) at (0,1) {};
        \node (02) at (0,2) {$\vdots$};
        \node[bd] (03) at (0,3) {};
        \node[bd] (41) at (4,1) {};
        \node (42) at (4,2) {$\vdots$};
        \node[bd] (43) at (4,3) {};
        \draw (00)--(10)--(20)--(30)--(40)--(41)--(42)--(43)--(44)--(34)--(24)--(14)--(04)--(03)--(02)--(01)--(00);
        \draw [decorate,decoration={brace,amplitude=5pt},xshift=0pt,yshift=0pt] (4.1,-0.3)--(-0.1,-0.3) node [black,midway,yshift=-0.5cm] {$n$ edges};
        \draw [decorate,decoration={brace,amplitude=5pt},xshift=0pt,yshift=0pt] (-0.3,-0.1)--(-0.3,4.1) node [black,midway,xshift=-0.5cm,rotate=90] {$k$ edges};
        \node[bd] (11) at (1,1) {};
        \node[bd] (13) at (1,3) {};
        \node[bd] (31) at (3,1) {};
        \node[bd] (33) at (3,3) {};
        \def\x{0.3};
        \draw (00)--(11)--(1+\x,1+\x) (01)--(0+\x,1+\x) (03)--(14) (10)--(1+\x,0+\x) (1-\x,3-\x)--(13)--(1+\x,3+\x) (30)--(41) (3-\x,1-\x)--(31)--(3+\x,1+\x) (3-\x,3-\x)--(33)--(44) (3-\x,4-\x)--(34) (4-\x,3-\x)--(43);
    \end{tikzpicture}}}\;.
\end{split}
\end{equation}

The latter two do not have a Higgs branch, therefore we cannot exploit them in our analysis to recover the stratification of Higgs branch of the 5d SCFT of interest.

\subsection{Dynamical deformations for $\mathcal T_{n,k}$ theories}\label{sec: normalizable def}

Before we dive in the detailed discussion of the $\mathcal T_{3,3}$ theory in this Section we discuss dynamical deformations for the toric threefolds above. We adopt a geometric perspective focusing on obtaining an explicit expression for the dynamical deformations of the threefold. Algebraically, this involves the study of complete intersections (or hypersurface) equations. We structure the analysis as follows:
\begin{enumerate}
    \item Start from a singular geometry of type:
    \begin{equation}\label{general molecule}
    \begin{cases}
        xy=z^n\\
        uv = F(x,y,z) \\
    \end{cases},
\end{equation}
i.e.\ a $\mathbb{C}^*$ fibration on a Du Val singularity of type $A_{n-1}$. The $\mathbb{C}^*$-action is:
\begin{equation}\label{C star}
    (u,v,x,y,z) \quad \longrightarrow \quad (\lambda u,\lambda^{-1}v,x,y,z).
\end{equation}
In particular, the box diagrams such as \eqref{eq:boxdiagram} are described by the threefold:
 \begin{equation}\label{box diagram eq}
    \begin{cases}
        xy=z^n,\\
        uv = z^k. \\
    \end{cases}
\end{equation}
    \item It is well known that the 5d Higgs branch is related to dynamical deformations of the singular geometry. The threefolds contain singular non-compact complex lines: in such cases deformation theory is not rigorously defined, as the Milnor ring is infinite dimensional. Nonetheless, heavily relying on the tools developed of \cite{Collinucci:2020jqd}, we are able to identify the (finite number of) genuine dynamical Higgs branch deformations.\\
    Let us summarize their argument here, applying it to our general setting: the procedure kicks off by performing a specific partial resolution of the singular threefold in a suitable toric ambient space. Such operation is a standard resolution of Du Val singularity in the first equation in \eqref{general molecule}, subject to the base change imposed by the second equation. For the box diagrams \eqref{box diagram eq}, we can choose either equation up to renaming of the variables, due to the evident symmetry, and interpret the leftover equation as a base change, defining a $\mathbb C^*$ fibration over the Du Val singularity. For concreteness, consider the crepant resolution of the first equation of \eqref{general molecule}: it inflates $n-1$ $\mathbb{P}^1$'s, intersecting like the $A_{n-1}$ Dynkin diagram. On top of each of these $\mathbb{P}^1$'s singularities of type $A$ are still present. This data allows us to construct a Lagrangian quiver of shape $\mathfrak{g}$ with $\mathfrak{su}$ nodes dictated by the remaining singularities of type $A$. The quiver hence corresponds to low-energy phase of the initial 5d SCFT point. In this setup, the gauge couplings are identified with the K\"ahler volumes of the $\mathbb P^1$'s, and are mass deformations of the SCFT. The blow-down maps corresponding to this partial resolution take the generic form:
    \begin{equation}\label{partial resolution}
    \begin{split}
     &   x = f_1(e_j,z_1,z_2) \\
     &   y = f_2(e_j,z_1,z_2) \\
     &   z = f_3(e_j,z_1,z_2) \\
    \end{split}\quad,
    \end{equation}
    where the $e_j,z_1,z_2$ are the homogeneous coordinates after the blowup, and the $f_i$ are polynomial functions. The $\mathbb C^*$-fibration is transparently visible from the second equation of \eqref{general molecule}. Namely, we can write it as a polynomial function of the blow-up coordinates:
    \begin{equation}\label{brane locus UV}
       uv = F(x,y,z)\big|_{x=x(e_j,z_1,z_2),y=y(e_j,z_1,z_2),z=z(e_j,z_1,z_2)} \equiv \Delta_{D6}(e_j,z_1,z_2).
    \end{equation}
    We  denote the degeneration locus of the $\mathbb C^*$-fiber as $\Delta_{D6}$, since we can use the compact $\mathbb S^1 \subset \mathbb C^*$ as the M-theory circle to take a Type IIA limit, obtaining \cite{Sen:1997kz,Closset_2019} stacks of D6 branes wrapped on each irreducible components of $\Delta_{D6}$.
    Hence,  \eqref{brane locus UV} represents the D6 brane-locus in Type IIA. 
    More specifically, the coordinates $d_j$ and their multiplicity in the factorized brane locus encode the number of D6-branes wrapping each of the exceptional $\mathbb{P}^1$'s of the $A_{n-1}$ algebra.
\item    From the perspective of the Type IIA reduction \eqref{brane locus UV}, \cite{Collinucci:2020jqd} proves that the correct Higgs branch deformation counting is produced if one discards deformations that act on a non-compact stack of D6-branes, as they require infinite energy and are thus non-dynamical. As far as the brane locus is concerned, this is reflected in requiring that one adds to equation \eqref{brane locus UV} only deformations that satisfy the following requirements:
    \begin{itemize}
        \item The deformations $\lambda_{k}$ must be invariant under \textit{all} the GLSM $\mathbb{C}^*$-actions enjoyed by the the GLSM presentation of the resolved $A_{n-1}$ singularity. This can be equivalently restated as restricting the deformations to be a polynomial function of $x,y,z$.
        Thus:
        \begin{equation}\label{deformation}
            \lambda_k = \lambda_k(x,y,z).
        \end{equation}
        \item Inserting the blow-down map into \eqref{deformation}, the allowed deformations must  \textit{not} be proportional to the \textit{compact} brane locus \eqref{brane locus UV}. Physically, if the deformation \textit{were} proportional to the compact brane locus, it would be acting on some non-compact stack of D6-branes, and therefore it should not be taken into account. Hence, this second requirement imposes that the low-energy quiver produced adding the deformation \eqref{deformation} to the threefold contains \textit{at least} one gauge node with smaller rank than the corresponding node in the quiver produced by \eqref{general molecule}.
        \end{itemize}
        \item All in all, one obtains the deformed Calabi-Yau threefold geometry as:
        \begin{equation}\label{3fold deformation}
        \begin{cases}
        P_{\mathfrak{g}}(x,y,z)= 0\\
        uv = F(x,y,z) + \sum_k c_k\lambda_k(x,y,z) \\
    \end{cases},
\end{equation}
with $c_k$ complex coefficients and $k = 1,\ldots,\nu$ the number of dynamical complex deformations. Notice that the geometries in \eqref{3fold deformation} are a special case of the bifundamental 5d conformal matter theories studied in \cite{Bourget:2026ono}, where the explicit crepant resolution and the algorithm to extract the low-energy quiver description has been examined in full details, also for cases involving $D$ an $E$-type singularities.\\
\indent The total Higgs branch dimension in the UV can be neatly computed as:
\begin{equation}\label{UV HB}
    \text{dim}_{\mathbb{H}}\mathrm{HB}_{UV}= \nu + \text{dim}(A_{n-1}).
\end{equation}
Notice that expression \eqref{UV HB}  coincides with the Higgs branch dimension that can be computed from the low-energy quiver phase furnished by the partial resolution in \eqref{partial resolution}, provided that
\begin{equation}\label{nu}
    \nu = n_H-n_V -\text{rank}(A_{n-1}).
\end{equation}
The validity of \eqref{nu} can be  checked for all cases at hand. \eqref{nu} also takes into account the additional deformations that are visible given that the electric quivers for \eqref{general molecule} have $\mathfrak{su}$ nodes, in lieu of $\mathfrak{u}$ ones.
Furthermore, as many deformations as the roots of $A_{n-1}$ must be added to $\nu$ in \eqref{nu} in order to obtain the full UV Higgs branch dimension \eqref{UV HB}, as our procedure has selected a specific partial resolution that freezes precisely that amount of wrapped M2-brane modes. For more details on this aspect, see the analysis in Section 3.2 of \cite{DeMarco:2025ugw}.

Crucially, tuning one of the coefficients $c_k$ in the deformed equation \eqref{3fold deformation} to be non-vanishing (with all the others set to zero) triggers an Higgs branch RG flow, ending up in the 5d SCFT corresponding to the deformation $\lambda_k$:
\begin{equation}\label{transition geometry}
        \begin{cases}
        xy=z^n\\
     uv=  F(x,y,z)
    \end{cases} \quad \xrightarrow{Higgsing} \quad \begin{cases}
        xy=z^n\\
        uv =F(x,y,z)+ \lambda_k(x,y,z) \\
    \end{cases}.
\end{equation}
 Of course, one can turn on multiple deformation parameters: these will in general turn on a RG flow that lands on different leaves, depending on the specific tunings of the coefficients.
We emphasize that this candidate RG flow along the Higgs branch is visible from the deformed geometry. 

It is now natural to ponder whether \textit{all} the transitions \eqref{transition geometry} can be observed from alternative perspectives, such as via the quiver subtraction technology \cite{Cabrera:2018ann,Bourget:2019aer}.

\end{enumerate}

\subsection{Higgs branch of a gauge theory phase of the $\mathcal T_{3,3}$ model}\label{sec: HB quivers T33}

Here we focus on the case with $n=k=3$:
\begin{equation}\label{CY3 T33}
\mathcal{T}_{3,3}: \quad \begin{cases}
   xy = z^3 \\
  uv= z^3\\
 \end{cases}.
\end{equation}
We can perform a full resolution of the Du Val singularity in the first equation, that corresponds to the partial resolution of the threefold, gaining access in this way to the 5d quiver phase  $\hQ{3}{3}$. Of course, performing a resolution of the second equation reproduces $\vQ{3}{3}$. By means of this partial resolutions we move in the extended Coulomb branch of the SCFT, resulting in gauge theory phases. The latter have their own Higgs branches which we characterize here. The partial resolution of the first equation amounts to introducing the toric homogeneous coordinates 
\begin{equation}\label{toric actions A2}
\renewcommand{\arraystretch}{1}
\begin{array}{c|cccc}
&z_1 & e_1 & e_2 & z_2  \\
\hline
 \mathbb C^*_1&1 & -2 & 1 & 0  \\
  \mathbb C^*_2&0 & 1 & -2 & 1  \\
\end{array}
\end{equation}
and the partial resolution map
\begin{equation}
\label{eq:blowdownA2}
    x = e_1 e_2^2 z_2^3, \quad y = z_1^3 e_1^2 e_2, \quad z = z_1 e_1 e_2 z_2,
\end{equation}
that automatically solves the first equation of \eqref{CY3 T33}. 
The (homogeneous) coordinates on the partial resolution are hence $(z_1, e_1, e_2, z_2, u,v)$, subject to the second equation of \eqref{CY3 T33}, and identified via the $\mathbb C^*_1$ and $\mathbb C^*_2$ actions of \eqref{toric actions A2}. Let us now count the dynamical complex deformations of this partially resolved phase, that have to be identified with the HB deformations of the electric gauge phase $\hQ{3}{3}$, that we reproduce here for convenience: 
\begin{equation}
        \scalebox{0.65}{\begin{tikzpicture}
    \draw[thick] (-0.7, -0.7) rectangle (0.7, 0.7);
    \node at (0, 0) {\small$3$};
    
    \draw[thick] (0.7, 0)--(1.3, 0);
    
    \draw[thick] (2, 0) circle (0.65);
    \node at (2, 0) {\small$3$};
    
    \draw[thick] (2.7, 0)--(3.3, 0);
    
    \draw[thick] (4, 0) circle (0.65);
    \node at (4, 0) {\small$3$};
    
    \draw[thick] (4.7, 0)--(5.3, 0);
    
    \draw[thick] (5.3, -0.7) rectangle (6.7, 0.7);
    \node at (6, 0) {\small$3$};
\end{tikzpicture}}
\end{equation}
Given this electric quiver the corresponding Higgs branch Hasse diagram is obtained via quiver subtraction. The unitary version of this quiver, and its unframed version, are
\begin{equation}
\label{eq:A2(1^3)ElQuivUnit}
    \mathsf{Q}_e^{\mathrm{unitary}}=\quad\raisebox{-.5\height}{\scalebox{0.7}{\begin{tikzpicture}
                \node[gaugeBig,label=below:{$3$}] (1) at (2,0) {};
                \node[gaugeBig,label=below:{$3$}] (2) at (4,0) {};
                \node[flavourSU,label=below:{$3$}] (1f) at (0,0) {};
                \node[flavourSU,label=below:{$3$}] (2f) at (6,0) {};
                \draw (1f)--(1)--(2)--(2f);
            \end{tikzpicture}}}\quad=\quad\raisebox{-.5\height}{\scalebox{0.7}{\begin{tikzpicture}
                \node[gaugeBig,label=below:{$3$}] (1) at (2,0) {};
                \node[gaugeBig,label=below:{$3$}] (2) at (4,0) {};
                \node[gaugeBig,label=above:{$1$}] (3) at (3,1.5) {};
                \draw (3)--(1)--(2)--(3);
                \draw[transform canvas={xshift=-3pt,yshift=3pt}] (1)--(3);
                \draw[transform canvas={xshift=3pt,yshift=-3pt}] (1)--(3);
                \draw[transform canvas={xshift=3pt,yshift=3pt}] (2)--(3);
                \draw[transform canvas={xshift=-3pt,yshift=-3pt}] (2)--(3);
            \end{tikzpicture}}}\quad.
\end{equation}
Performing quiver subtraction on the unframed unitary quiver yields
\begin{equation}
\label{eq:electricsubtractiona2}
    \scalebox{0.7}{\begin{tikzpicture}
        \node (a) at (0,0) {$\raisebox{-.5\height}{\scalebox{0.7}{\begin{tikzpicture}
                \node[gaugeBig,label=below:{$3$}] (1) at (2,0) {};
                \node[gaugeBig,label=below:{$3$}] (2) at (4,0) {};
                \node[gaugeBig,label=above:{$1$}] (3) at (3,1.5) {};
                \draw (3)--(1)--(2)--(3);
                \draw[transform canvas={xshift=-3pt,yshift=3pt}] (1)--(3);
                \draw[transform canvas={xshift=3pt,yshift=-3pt}] (1)--(3);
                \draw[transform canvas={xshift=3pt,yshift=3pt}] (2)--(3);
                \draw[transform canvas={xshift=-3pt,yshift=-3pt}] (2)--(3);
            \end{tikzpicture}}}$};
        \node (b) at (-2,-4) {$\raisebox{-.5\height}{\scalebox{0.7}{\begin{tikzpicture}
                \node[gaugeBig,label=below:{$2$}] (1) at (2,0) {};
                \node[gaugeBig,label=below:{$3$}] (2) at (4,0) {};
                \node[gaugeBig,label=above:{$1$}] (3) at (3,1.5) {};
                \draw (3)--(1)--(2);
                \draw[transform canvas={xshift=1pt,yshift=1pt}] (2)--(3);
                \draw[transform canvas={xshift=3pt,yshift=3pt}] (2)--(3);
                \draw[transform canvas={xshift=-1pt,yshift=-1pt}] (2)--(3);
                \draw[transform canvas={xshift=-3pt,yshift=-3pt}] (2)--(3);
            \end{tikzpicture}}}$};
        \node (c) at (2,-4) {$\raisebox{-.5\height}{\scalebox{0.7}{\begin{tikzpicture}
                \node[gaugeBig,label=below:{$2$}] (1) at (2,0) {};
                \node[gaugeBig,label=below:{$3$}] (2) at (4,0) {};
                \node[gaugeBig,label=above:{$1$}] (3) at (3,1.5) {};
                \draw (1)--(2)--(3);
                \draw[transform canvas={xshift=-1pt,yshift=1pt}] (1)--(3);
                \draw[transform canvas={xshift=-3pt,yshift=3pt}] (1)--(3);
                \draw[transform canvas={xshift=1pt,yshift=-1pt}] (1)--(3);
                \draw[transform canvas={xshift=3pt,yshift=-3pt}] (1)--(3);
            \end{tikzpicture}}}$};
        \node (d) at (0,-8) {$\raisebox{-.5\height}{\scalebox{0.7}{\begin{tikzpicture}
                \node[gaugeBig,label=below:{$2$}] (1) at (2,0) {};
                \node[gaugeBig,label=below:{$2$}] (2) at (4,0) {};
                \node[gaugeBig,label=above:{$1$}] (3) at (3,1.5) {};
                \draw (1)--(2);
                \draw[transform canvas={xshift=-2pt,yshift=2pt}] (1)--(3);
                \draw[transform canvas={xshift=2pt,yshift=-2pt}] (1)--(3);
                \draw[transform canvas={xshift=2pt,yshift=2pt}] (2)--(3);
                \draw[transform canvas={xshift=-2pt,yshift=-2pt}] (2)--(3);
            \end{tikzpicture}}}$};
        \node (e) at (-2,-12) {$\raisebox{-.5\height}{\scalebox{0.7}{\begin{tikzpicture}
                \node[gaugeBig,label=below:{$1$}] (1) at (2,0) {};
                \node[gaugeBig,label=below:{$2$}] (2) at (4,0) {};
                \node[gaugeBig,label=above:{$1$}] (3) at (3,1.5) {};
                \draw (1)--(2)--(3);
                \draw[transform canvas={xshift=3pt,yshift=3pt}] (2)--(3);
                \draw[transform canvas={xshift=-3pt,yshift=-3pt}] (2)--(3);
            \end{tikzpicture}}}$};
        \node (f) at (2,-12) {$\raisebox{-.5\height}{\scalebox{0.7}{\begin{tikzpicture}
                \node[gaugeBig,label=below:{$2$}] (1) at (2,0) {};
                \node[gaugeBig,label=below:{$1$}] (2) at (4,0) {};
                \node[gaugeBig,label=above:{$1$}] (3) at (3,1.5) {};
                \draw (3)--(1)--(2);
                \draw[transform canvas={xshift=-3pt,yshift=3pt}] (1)--(3);
                \draw[transform canvas={xshift=3pt,yshift=-3pt}] (1)--(3);
            \end{tikzpicture}}}$};
        \node (g) at (0,-16) {$\raisebox{-.5\height}{\scalebox{0.7}{\begin{tikzpicture}
                \node[gaugeBig,label=below:{$1$}] (1) at (2,0) {};
                \node[gaugeBig,label=below:{$1$}] (2) at (4,0) {};
                \node[gaugeBig,label=above:{$1$}] (3) at (3,1.5) {};
                \draw (3)--(1)--(2)--(3);
            \end{tikzpicture}}}$};
        \node (h) at (0,-20) {$\raisebox{-.5\height}{\scalebox{0.7}{\begin{tikzpicture}
                \node[gaugeBig,label=above:{$1$}] (3) at (3,1.5) {};
            \end{tikzpicture}}}$};
        \draw[->] (a)--(b);
        \node at (-1.5,-2) {$-\raisebox{-.5\height}{\scalebox{0.4}{\begin{tikzpicture}
                \node[gaugeBig,label=below:{$1$}] (1) at (2,0) {};
                \node[gaugeBig,label=above:{$1$}] (3) at (3,1.5) {};
                \draw (3)--(1);
                \draw[transform canvas={xshift=-3pt,yshift=3pt}] (1)--(3);
                \draw[transform canvas={xshift=3pt,yshift=-3pt}] (1)--(3);
            \end{tikzpicture}}}$};
        \draw[->] (a)--(c);
        \node at (1.5,-2) {$-\raisebox{-.5\height}{\scalebox{0.4}{\begin{tikzpicture}
                \node[gaugeBig,label=below:{$1$}] (2) at (4,0) {};
                \node[gaugeBig,label=above:{$1$}] (3) at (3,1.5) {};
                \draw (3)--(2);
                \draw[transform canvas={xshift=3pt,yshift=3pt}] (2)--(3);
                \draw[transform canvas={xshift=-3pt,yshift=-3pt}] (2)--(3);
            \end{tikzpicture}}}$};
        \draw[->] (b)--(d);
        \node at (-1.5,-6) {$-\raisebox{-.5\height}{\scalebox{0.4}{\begin{tikzpicture}
                \node[gaugeBig,label=below:{$1$}] (2) at (4,0) {};
                \node[gaugeBig,label=above:{$1$}] (3) at (3,1.5) {};
                \draw[transform canvas={xshift=3pt,yshift=3pt}] (2)--(3);
                \draw[transform canvas={xshift=1pt,yshift=1pt}] (2)--(3);
                \draw[transform canvas={xshift=-1pt,yshift=-1pt}] (2)--(3);
                \draw[transform canvas={xshift=-3pt,yshift=-3pt}] (2)--(3);
            \end{tikzpicture}}}$};
        \draw[->] (c)--(d);
        \node at (1.6,-6) {$-\raisebox{-.5\height}{\scalebox{0.4}{\begin{tikzpicture}
                \node[gaugeBig,label=below:{$1$}] (1) at (2,0) {};
                \node[gaugeBig,label=above:{$1$}] (3) at (3,1.5) {};
                \draw[transform canvas={xshift=-3pt,yshift=3pt}] (1)--(3);
                \draw[transform canvas={xshift=-1pt,yshift=1pt}] (1)--(3);
                \draw[transform canvas={xshift=1pt,yshift=-1pt}] (1)--(3);
                \draw[transform canvas={xshift=3pt,yshift=-3pt}] (1)--(3);
            \end{tikzpicture}}}$};
        \draw[->] (d)--(e);
        \node at (-1.5,-10) {$-\raisebox{-.5\height}{\scalebox{0.4}{\begin{tikzpicture}
                \node[gaugeBig,label=below:{$1$}] (1) at (2,0) {};
                \node[gaugeBig,label=above:{$1$}] (3) at (3,1.5) {};
                \draw[transform canvas={xshift=-2pt,yshift=2pt}] (1)--(3);
                \draw[transform canvas={xshift=2pt,yshift=-2pt}] (1)--(3);
            \end{tikzpicture}}}$};
        \draw[->] (d)--(f);
        \node at (1.5,-10) {$-\raisebox{-.5\height}{\scalebox{0.4}{\begin{tikzpicture}
                \node[gaugeBig,label=below:{$1$}] (2) at (4,0) {};
                \node[gaugeBig,label=above:{$1$}] (3) at (3,1.5) {};
                \draw[transform canvas={xshift=2pt,yshift=2pt}] (2)--(3);
                \draw[transform canvas={xshift=-2pt,yshift=-2pt}] (2)--(3);
            \end{tikzpicture}}}$};
        \draw[->] (e)--(g);
        \node at (-1.5,-14) {$-\raisebox{-.5\height}{\scalebox{0.4}{\begin{tikzpicture}
                \node[gaugeBig,label=below:{$1$}] (2) at (4,0) {};
                \node[gaugeBig,label=above:{$1$}] (3) at (3,1.5) {};
                \draw (3)--(2);
                \draw[transform canvas={xshift=3pt,yshift=3pt}] (2)--(3);
                \draw[transform canvas={xshift=-3pt,yshift=-3pt}] (2)--(3);
            \end{tikzpicture}}}$};
        \draw[->] (f)--(g);
        \node at (1.6,-14) {$-\raisebox{-.5\height}{\scalebox{0.4}{\begin{tikzpicture}
                \node[gaugeBig,label=below:{$1$}] (1) at (2,0) {};
                \node[gaugeBig,label=above:{$1$}] (3) at (3,1.5) {};
                \draw (3)--(1);
                \draw[transform canvas={xshift=-3pt,yshift=3pt}] (1)--(3);
                \draw[transform canvas={xshift=3pt,yshift=-3pt}] (1)--(3);
            \end{tikzpicture}}}$};
        \draw[->] (g)--(h);
        \node at (-0.8,-18.2) {$-\raisebox{-.5\height}{\scalebox{0.4}{\begin{tikzpicture}
                \node[gaugeBig,label=below:{$1$}] (1) at (2,0) {};
                \node[gaugeBig,label=below:{$1$}] (2) at (4,0) {};
                \node[gaugeBig,label=above:{$1$}] (3) at (3,1.5) {};
                \draw (3)--(1)--(2)--(3);
            \end{tikzpicture}}}$};
    \end{tikzpicture}}\;.
\end{equation}
Back to geometry, in terms of the homogeneous coordinates, the second equation of \eqref{CY3 T33} reads
\begin{equation}
\label{eq:A2branelocus}
    uv= z_1^3 e_1^3 e_2^3 z_2^3 = \Delta_{D6}(z_1,e_1,e_2,z_2),
\end{equation}
where we refer to the r.h.s.\ of \eqref{eq:A2branelocus} as ``brane locus'', as explained in \eqref{brane locus UV}. In this case, performing a reduction to IIA harnessing the circle contained in the $\mathbb{C}^*$-action dualizes the M-theory setup to 4 stacks of 3 D6 branes, each stack wrapped on one of the (compact and non-compact) 2-cycles of the resolved $A_2$ singularity, reproducing in this way $\hQ{3}{3}$. The 2-cycles of the resolved $A_2$ singularity are given  by the zero-loci of the ideals:
\begin{equation}
\begin{array}{l}
    \mathcal{C}_1 =(e_1)\\
    \mathcal{C}_2 =(e_2) \\
    \mathcal{C}_{l} =(z_1) \\
    \mathcal{C}_{r} =(z_2) 
\end{array} \quad\quad
    \scalebox{0.65}{\begin{tikzpicture}
    \draw[thick] (-0.7, -0.7) rectangle (0.7, 0.7);
    \node at (0, 0) {\small$\mathcal{C}_l$};
    
    \draw[thick] (0.7, 0)--(1.3, 0);
    
    \draw[thick] (2, 0) circle (0.65);
    \node at (2, 0) {\small$\mathcal{C}_1$};
    
    \draw[thick] (2.7, 0)--(3.3, 0);
    
    \draw[thick] (4, 0) circle (0.65);
    \node at (4, 0) {\small$\mathcal{C}_2$};
    
    \draw[thick] (4.7, 0)--(5.3, 0);
    
    \draw[thick] (5.3, -0.7) rectangle (6.7, 0.7);
    \node at (6, 0) {\small$\mathcal{C}_r$};
\end{tikzpicture}}
\end{equation}
where we denote with rectangles the non-compact curves wrapped by the D6s.

\medskip

As we have reviewed in Section \ref{sec: normalizable def}, the logic of \cite{Collinucci:2020jqd} shows that dynamical deformations in the partially resolved phase are those that \textit{are not} proportional to the compact brane locus \cite{Collinucci:2020jqd}: deformations proportional to the compact brane locus modify the position of the (infinitely massive) non-compact D6 branes, and are hence non-dynamical. In this case, the compact brane locus is (the zero locus of) $e_1^3 e_2^3$, and the \textit{dynamical} deformations, as it can be checked via the explicit expression of the blowdown map \eqref{eq:blowdownA2}, deform \eqref{CY3 T33} to: 
\begin{equation}
    \begin{cases}\label{A2 molecule deformed}
    xy= z^3 \\
   uv = z^{3} + c_1 + c_2 x + c_3 y+  c_4 z + c_5 x^2 + c_6 y^2 + c_7 z^2 + c_8 xz + c_9 yz. \\
 \end{cases},
\end{equation}
with $c_j$, with  $j = 1,..., \nu = 9$ being the deformation coefficients. The full HB, after sending the gauge couplings of $\hQ{3}{3}$ to infinity, has dimension
\begin{equation}\label{UV HB T33}
\text{dim}_{\mathbb{H}}\mathrm{HB}\left(\mathcal{T}_{3,3}\right) = \nu + \text{dim}(\mathfrak{g})  = 9+\text{dim}(A_2) = 17,
\end{equation}
which reproduces the result that can be obtained directly from the quiver description.\\
Let us now match the possible deformations (and the corresponding Higgsings) of \eqref{A2 molecule deformed} to \eqref{eq:electricsubtractiona2}. 
First, we show the explicit expressions of the deformations: 
\begin{eqnarray}
\label{eq:monomiallist}
    x =  e_1 e_2^2 z_2^3, \quad y = z_1^3 e_1^2 e_2, \quad  z = z_1 e_1 e_2 z_2, \nonumber \\
    x^2 =  e_1^2 e_2^4 z_2^6, \quad y^2 = z_1^6 e_1^4 e_2^2, \quad  z^2 = z_1^2 e_1^2 e_2^2 z_2^2, \nonumber \\
    xz =  z_1 e_1^2 e_2^3 z_2^4, \qquad yz = z_1^4 e_1^3 e_2^2 z_1. \nonumber \\
\end{eqnarray}

One then proceeds as follows: switching on the complex deformations in \eqref{A2 molecule deformed} one is also deforming the initial D6-brane locus, which re-organizes into the following framed quiver, where the occupation numbers depend on the choice of coefficients $c_i$:

\begin{equation}
\label{eq:listoframedquivers}
        \scalebox{0.65}{\begin{tikzpicture}
    \draw[thick] (-0.7, -0.7) rectangle (0.7, 0.7);
    \node at (0, 0) {\small$m_1$};
    
    \draw[thick] (0.7, 0)--(1.3, 0);
    
    \draw[thick] (2, 0) circle (0.65);
    \node at (2, 0) {\small$m_2$};
    
    \draw[thick] (2.7, 0)--(3.3, 0);
    
    \draw[thick] (4, 0) circle (0.65);
    \node at (4, 0) {\small$m_3$};
    
    \draw[thick] (4.7, 0)--(5.3, 0);
    
    \draw[thick] (5.3, -0.7) rectangle (6.7, 0.7);
    \node at (6, 0) {\small$m_4$};
\end{tikzpicture}}
\end{equation}
Employing the explicit expressions of the deformations \eqref{eq:monomiallist}, \textit{we can then match the Hasse diagram \eqref{eq:electricsubtractiona2} re-phrasing each of the quivers corresponding to the deformations \eqref{eq:listoframedquivers} to the unframed notation}. We can pictorially sum up the result by reproducing the Hasse diagram obtained by quiver subtraction appropriately tuning the coefficients in \eqref{A2 molecule deformed}.\\
\indent Next to each point we write down which monomial deformations \textit{must} be turned on in order for the Higgs Branch RG flow to land on that specific leaf. If two or more monomials are separated by a comma, it means that \textit{at least} one of them must be present so that the leaf is reproduced. On a given leaf, any monomial that appears in leaves \textit{below} it can also appear as a deformation, without modifying the brane locus.
\begin{equation}
\label{eq:electricsubtractiona2monomials}
    \scalebox{0.7}{\begin{tikzpicture}
        \node (a) at (0,-12) {$(1)$};
        \node (b) at (0,-9) {$(z)$};
        \node (c) at (-2,-6) {$(x)$};
        \node (d) at (2,-6) {$(y)$};
        \node (e) at (0,-3) {
        \small$(z^2)$};
        \node (f) at (-2,0) {$(x^2,xz)$};
        \node (g) at (2,0) {$(y^2,yz)$};
        \node (h) at (0,3) {$(\emptyset)$};

        \draw[-] (a)--(b);
        \draw[-] (b)--(c);
        \draw[-] (b)--(d);
        \draw[-] (c)--(e);
        \draw[-] (d)--(e);
        \draw[-] (e)--(f);
        \draw[-] (e)--(g);
        \draw[-] (f)--(h);
        \draw[-] (g)--(h);
    \end{tikzpicture}}
\end{equation}

Let us explicitly show how one reconstructs the quiver from the monomial deformations. In general, they might encode the recombination of compact and non-compact D6-branes into non-compact stacks of bound D6-branes. E.g.\ considering the leaf labelled by $(x^2,xz)$ in Figure \ref{eq:electricsubtractiona2monomials} we get:
\begin{equation}\label{subtle def}
    \begin{cases}
        xy=z^3\\
uv=z^3+c_5x^2+c_8xz\big|_{x=x(z_i,e_i),y=y(z_i,e_i),z=z(z_i,e_i)} = e_1^2 e_2^3 z_2^3 \left(e_1 z_1^3+c_5 e_2 z_2^3+c_8 z_1 z_2\right)\\
    \end{cases}.
\end{equation}
Hence the brane locus reads:
\begin{equation}
\label{eq:example33branelocus}
    \Delta_{D6}(z_1,e_1,e_2,z_2) = e_1^2 e_2^3 z_2^3 \left(e_1 z_1^3+c_5 e_2 z_2^3+c_8 z_1 z_2\right),
\end{equation}
where one can clearly spot the recombined flavor brane in brackets. The recombined brane in \eqref{eq:example33branelocus}  intersects the curves $e_1 = 0$ and $e_2 =0$ each with intersection number one. Hence, the brane-locus  \eqref{eq:example33branelocus} corresponds to the following special unitary quiver: 
\begin{equation}
        \scalebox{0.65}{\begin{tikzpicture}
    \draw[thick] (-0.7, -0.7) rectangle (0.7, 0.7);
    \node at (0, 0) {\small$1$};
    
    \draw[thick] (0.7, 0)--(1.3, 0);
    
    \draw[thick] (2, 0) circle (0.65);
    \node at (2, 0) {\small$2$};
    
    \draw[thick] (2.7, 0)--(3.3, 0);
    
    \draw[thick] (4, 0) circle (0.65);
    \node at (4, 0) {\small$3$};
    
    \draw[thick] (4.7, 0)--(5.3, 0);
    
    \draw[thick] (5.3, -0.7) rectangle (6.7, 0.7);
    \node at (6, 0) {\small$4$};
\end{tikzpicture}}
\end{equation}
This transition can also be readily interpreted in terms of quiver subtraction, showing perfect agreement between the geometric and the Hasse diagram perspective.\\
\indent This reasoning beautifully connects with the recent literature on T-branes and GTPs \cite{Bourget:2023wlb,Alexeev:2024bko}, on which we will delve into more details in Section \ref{sec:white dots}.\\

\indent Up to this stage we have focused on normalizable deformations of the toric CY3 \eqref{CY3 T33} that modify the first equation, and keep the second equation unchanged. This reproduces the dynamical deformations of $\hQ{3}{3}$. Naturally, one could as easily consider deformations of the second equation, leaving the first unchanged. This reproduces the dynamical deformations of $\vQ{3}{3}$. Namely, one obtains a Hasse diagram identical to Figure \ref{eq:electricsubtractiona2monomials}, with swapped variables $x \leftrightarrow u, y \leftrightarrow v$:
\begin{equation}
\label{eq:electricsubtractiona2monomialsv2}
     \scalebox{0.7}{\begin{tikzpicture}
        \node (a) at (0,-12) {$(1)$};
        \node (b) at (0,-9) {$(z)$};
        \node (c) at (-2,-6) {$(u)$};
        \node (d) at (2,-6) {$(v)$};
        \node (e) at (0,-3) {
        \small$(z^2)$};
        \node (f) at (-2,0) {$(u^2,uz)$};
        \node (g) at (2,0) {$(v^2,vz)$};
        \node (h) at (0,3) {$(\emptyset)$};

        \draw[-] (a)--(b);
        \draw[-] (b)--(c);
        \draw[-] (b)--(d);
        \draw[-] (c)--(e);
        \draw[-] (d)--(e);
        \draw[-] (e)--(f);
        \draw[-] (e)--(g);
        \draw[-] (f)--(h);
        \draw[-] (g)--(h);
    \end{tikzpicture}}
\end{equation}

In the next Section, we explore how we can combine the perspectives gained from $\hQ{3}{3}$ and $\vQ{3}{3}$ in order to explore the full Higgs Branch of the 5d SCFT engineered by \eqref{CY3 T33}.

\subsection{Geometric Hasse diagram of $\mathcal{T}_{3,3}$}\label{sec:UV HB box}
In order to compute the Higgs branch dimension of the UV fixed point corresponding to the toric Calabi-Yau in \eqref{CY3 T33}, one should naturally take into account the dynamical deformations in both the equations that define said CY3.
In general, we can combine the two perspectives of the previous section to obtain the fully deformed phase of the initial SCFT.
As we have remarked earlier, in order to manifestly compute these deformations one has to focus on one equation, and consider the remaining one as a $\mathbb{C}^*$-fibration over it, thus openly showing the Type IIA reduction and the D6-brane interpretation. Retracing these steps for both equations yields two different Calabi-Yau's, corresponding to $\hQ{3}{3}$ and $\vQ{3}{3}$:
\begin{equation}\label{both deformed}
\begin{split}
  &  \begin{cases}
        xy=z^3\\
        uv=z^3+ c_1 + c_2 x + c_3 y+  c_4 z + c_5 x^2 + c_6 y^2 + c_7 z^2 + c_8 xz + c_9 yz\\
    \end{cases}\\
   & \begin{cases}
        xy=z^3+ k_1 + k_2 u+ + k_3 v+  k_4 z + k_5 u^2 + k_6 v^2 + k_7 z^2 + k_8 uz + k_9 vz\\
        uv=z^3\\
    \end{cases}
    \end{split}
\end{equation}
We would like to write down all the deformations into an expression for the same Calabi-Yau. To this end, and to ensure the correct counting of the UV Higgs branch dimension, we must discard deformation parameters in \eqref{both deformed} that can be seen just as a change of coordinates. In the case of \eqref{both deformed}, we start by turning on all the deformations at the same time: 
\begin{equation}\label{def CY3}
     \begin{cases}
        xy=z^3+ k_1+  k_2 u+ + k_3 v+  k_4 z + k_5 u^2 + k_6 v^2 + k_7 z^2 + k_8 uz + k_9 vz\\
        uv=z^3+ c_1 + c_2 x + c_3 y+  c_4 z + c_5 x^2 + c_6 y^2 + c_7 z^2 + c_8 xz + c_9 yz\\
    \end{cases}\\,
\end{equation}
Shifting $z$ by a constant and renaming the deformation parameters we can e.g.\ eliminate the deformation ``$k_1$'', thus obtaining the expression for the fully deformed CY3: 
\begin{equation}\label{def CY3 T33}
     \begin{cases}
        xy=z^3+ k_2 u+ + k_3 v+  k_4 z + k_5 u^2 + k_6 v^2 + k_7 z^2 + k_8 uz + k_9 vz\\
        uv=z^3+ c_1 + c_2 x + c_3 y+  c_4 z + c_5 x^2 + c_6 y^2 + c_7 z^2 + c_8 xz + c_9 yz\\
    \end{cases}\\.
\end{equation}
Notice that in total there are 17 independent dynamical deformation parameters, as expected from \eqref{UV HB T33}. This is a reassuring check that the deformed Calabi-Yau \eqref{def CY3 T33} captures the possible directions along the Higgs branch of the SCFT.\\

\indent We now show that expression \eqref{def CY3 T33} contains detailed information about the Higgs branch RG flow, and therefore allows to completely reconstruct the HB stratification in terms of the geometric Hasse diagram. As we have seen considering deformations of only one equation that defines the CY3 reproduces a part of the Hasse diagram that corresponds to a gauge theory phase.  In the geometric language, this can be seen from \eqref{def CY3 T33} turning off all the coefficients $k_i$ (or $c_i$, for the symmetric equivalent). The remaning deformations instead give rise to leaves that arise from the simultaneous deformation of both equations that define the box-diagram. Due to the hierarchy dictated by gauge theory phase, we can order also the resulting deformations accordingly. It follows that in \eqref{def CY3 T33} \textit{at least} one of the $c_i$ and one of the $k_i$ are non-vanishing. Following this reasoning, we can obrtain the full geometric Hasse diagram starting from the analysis of the gauge theory phase. The latter in facts is dictated directly by the structure of the deformations of the non-isolated singularity, which in facts is gives the ordering of the corresponding dynamical deformations. As usual, some of the leaves of the Hasse diagram are labelled with a list $(m_1,...,m_l)$ of monomials. This means that we have to turn the deformation associated with \textit{at least one} among the $m_i$ to flow to the corresponding leaf.  E.g., as explained around \eqref{subtle def}, we need to deform at least by $x^2$ or by $xz$, to flow to the leaf denoted by $(x^2,xz)$. Some leaves are instead denoted by the direct sum of many round brackets. The direct sum symbol $(a_i)\oplus(b_i)$ indicates that at least one monomial among the $a_i$ \textit{and} one monomial among the $b_i$ must appear.\footnote{As noticed right after \eqref{both deformed}, deforming both equations with the same $z^2$ term leaves the singularity unchanged. Consequently, when we write $(z^2) \oplus (z^2)$, this means that we must use two different coefficients $k_7 \neq c_7$ to deform the two equations of \eqref{CY3 T33}.} Any monomials appearing in leaves \textit{above} it can also appear, without modifying the theory. This gives the following geometric Hasse diagram:

\begin{equation}
\label{Hasse complete T33}
     \scalebox{0.7}{\begin{tikzpicture}
        \node (a) at (0,-14) {$(1)$};
        \node (b) at (-12,-9) {$(x)$};
        \node (c) at (12,-9) {$(u)$};
        \node (d) at (-9,-9) {$(y)$};
        \node (e) at (9,-9) {$(v)$};
        \node (f) at (0,-9) {$(z^2)\oplus (z^2)$};
        \node (g) at (-10.5,-6) {$(z^2)$};
        \node (h) at (10.5,-6) {$(z^2)$};
        \node (j) at (-7,-6) {$(x^2,xz)\oplus(v^2,vz)$};
        \node (k) at (-3,-6) {$(x^2,xz)\oplus(u^2,uz)$};
        \node (l) at (7,-6) {$(y^2,yz)\oplus(u^2,uz)$};
        \node (m) at (3,-6) {$(y^2,yz)\oplus(v^2,vz)$};
        \node (n) at (-9,-3) {$(x^2,xz)$};
        \node (o) at (9,-3) {$(u^2,uz)$};
        \node (p) at (-6,-3) {$(y^2,yz)$};
        \node (q) at (6,-3) {$(v^2,vz)$};
        \node (r) at (0,0) {$(\emptyset)$};
        \draw[red] (a)--(b);
        \draw[cyan] (a)--(c);
        \draw[red] (a)--(d);
        \draw[cyan] (a)--(e);
        \draw[-] (a)--(f);
        \draw[red] (b)--(g);
        \draw[red] (d)--(g);
        \draw[cyan] (c)--(h);
        \draw[cyan] (e)--(h);
        \draw[-] (f)--(j);
        \draw[-] (f)--(k);
        \draw[-] (f)--(l);
        \draw[-] (f)--(m);
        \draw[-] (f)--(g);
        \draw[-] (f)--(h);
        \draw[red] (g)--(n);
        \draw[red] (g)--(p);
        \draw[cyan] (h)--(o);
        \draw[cyan] (h)--(q);
        \draw[red] (n)--(r);
        \draw[cyan] (o)--(r);
        \draw[red] (p)--(r);
        \draw[cyan] (q)--(r);
        \draw[-] (j)--(n);
        \draw[-] (j)--(q);
        \draw[-] (k)--(n);
        \draw[-] (k)--(o);
        \draw[-] (m)--(p);
        \draw[-] (m)--(q);
        \draw[-] (l)--(p);
        \draw[-] (l)--(o);
    \end{tikzpicture}}
\end{equation}
where we used the red and blue colors to denote the deformation associated, respectively, with the first and second system of equations appearing in \eqref{both deformed}, and that can be read from the respective IIA limits.
We claim that this geometric Hasse diagram in \eqref{Hasse complete T33} captures the structure of the stratification of the SCFT Higgs branch. In the following section we show that this indeed is consistent and completes the bridge between the quiver subtraction algorithm and its geometric interpretation: every leaf on the Hasse diagram can be reached via a dynamical deformation, that can be clearly detected from the explicit expression of the CY3 that engineers the UV fixed point.\\
\indent Furthermore, diagrams such as Figure \ref{Hasse complete T33} offer us a way to inspect the relationship between Calabi-Yau threefolds with isolated singularities and ones with non-isolated ones. For the sake of clarity, consider the point labelled by $(z^2)\oplus(z^2)$. This is engineered by:
\begin{equation}\label{def CY3 z2z2}
     \begin{cases}
        xy=z^3 + k_7 z^2+\ldots\\
        uv=z^3+ c_7 z^2 +\ldots\\
    \end{cases}\\,
\end{equation}
where the "$+\ldots$" indicates that also any combination of the terms that appear in leaves \textit{above} it are allowed, and do not modify the fixed point. Let us set the "$+\ldots$" terms to zero in both equations. Then the threefold sports four lines of non-isolated singularities, each supporting a Du Val double point of type $A_1$, corresponding to a $2\times2$ toric box diagram:
\begin{equation}\label{def CY3 z2z2 non iso}
    \begin{cases}
     xy=z^3 + k_7 z^2\\
        uv=z^3+ c_7 z^2 \\   
    \end{cases} \quad \Longrightarrow \quad \text{Singularities: } \begin{cases}
        x = y = z = u = 0\\
        x = y = z = v = 0\\
        u = v = z = x = 0\\
        u = v = z = y = 0\\
    \end{cases}
\end{equation}
The theory engineered by \eqref{def CY3 z2z2 non iso} admits a low-energy quiver phase which is $SU(2)$ with 4 flavors. Hence, the flavor symmetry of the UV point has rank 5, and the UV Higgs branch dimension is 7.\\
\indent Consider now a different choice for the terms "$+\ldots$" appearing in \eqref{def CY3 z2z2}, namely:
\begin{equation}\label{def CY3 z2z2 iso}
    \begin{cases}
        xy=z^3+k_5 u^2 + k_6 v^2 + k_7 z^2 \\
        uv=z^3+ c_5 x^2 + c_6 y^2 + c_7 z^2 \\
    \end{cases} \quad \Longrightarrow \quad \text{Singularities: } x = y = z = u = v = 0.
\end{equation}
Strikingly, the threefold \eqref{def CY3 z2z2 iso} only has an isolated singularity at the origin, but corresponds to the same UV fixed point as \eqref{def CY3 z2z2 non iso}. Convincing support for this claim can be provided noticing that:
\begin{itemize}
    \item the flavor symmetry of \eqref{def CY3 z2z2 iso} has rank 5 (as can be checked computing e.g.\ $\text{dim}(H_2(\mathbb{Z})-\text{dim}(H_4(\mathbb{Z})$ via an explicit crepant resolution).
    \item The deformation theory of \eqref{def CY3 z2z2 iso} is well defined, since it only has an isolated singularity. The Milnor number is 9. Employing the fact that the flavor symmetry has rank 5, we find 7 normalizable deformations,\footnote{Recall that, for isolated singular threefolds with Milnor number $\mu$ and that engineer a theory with flavor group of rank $f$, one has \cite{Closset:2020scj}:
    $\text{dim}_{\mathbb{H}}(HB) = \frac{\mu+f}{2}$} in agreement with the expectation coming from the non-isolated presentation \eqref{def CY3 z2z2 non iso} of the same UV fixed point.
\end{itemize}

Of course, we can check the consistency of our construction by computing the Higgs branch dimension of a leaf in the Hasse diagram \eqref{Hasse complete T33}. This is done retracing exactly the same steps as in the previous discussion: one considers the deformed threefold corresponding to a specific leaf as a new starting threefold, and counts normalizable deformations that are not proportional to its brane locus. Naturally, this reasoning holds for \textit{any} leaf in the Hasse diagram. Take, for example, the leaf given by:
\begin{equation}\label{T33 leaf}
    \begin{cases}
        xy = z^3 \\
        uv = z^3 + yz\\
    \end{cases},
\end{equation}
which is obtained from \eqref{def CY3 T33} turning off all deformations parameters, and setting $c_9 = 1$. The complete intersection in \eqref{T33 leaf} is per-se a perfectly well-defined threefold: as such, it admits a corresponding deformation theory. Computing the normalizable deformations for \eqref{T33 leaf} yields the deformed threefold:
\begin{equation}\label{T33 leaf def}
    \begin{cases}
        xy = z^3 +k_2 u+k_3 v+k_4 z+k_5u^2+k_6 v^2 +k_7z^2+k_8uz+k_9 vz\\
        uv = z^3 + yz+c_1+c_2x+c_3y+c_4 z+c_5x^2+c_7z^2+c_8x z\\
    \end{cases}.
\end{equation}
In total, one finds 15 normalizable deformations: this is in agreement with the magnetic quiver perspective, depicted in Figure \ref{fig:full UV T33}.\\
\indent For the case with generic $n$ and $k$ in the expression for the threefold \eqref{box diagram eq}, one proceeds exactly in the same fashion, deforming both equations.  In total, the number of independent dynamical deformations is:
\begin{equation}
\#\text{ of dynamical deformations}(\mathcal{T}_{n,k}) = n^2+k^2-1.
\end{equation}
The Hasse diagram can be recovered studying the resolution of the leaves.

\medskip

We conclude this Section with the following crucial comment.  In the example involving 5-brane webs shaped as a rectangular box examined so far, the dimension of the UV Higgs branch could be recovered considering deformations of both the equations defining the CY3, analyzed one at a time. Combining the two Type IIA perspectives arising from the two equations yields a full description of the UV Hasse diagram. 
In Section \ref{sec: Tn theories}, we show an example where the Type IIA reduction is not sufficient to recover the full UV Higgs branch, as additional normalizable deformations can appear.

\subsection{Consistency Check from Magnetic Quiver} 
\begin{figure}
\centering
\begin{tikzpicture}[
    scale=0.7,
    line cap=round,
    line join=round,
    brane/.style={very thick},
    sevenbrane/.style={circle, draw=black, fill=white, very thick, inner sep=0pt, minimum size=5.5mm}
]


\node[sevenbrane] (L3) at (-4.0,0) {};
\node[sevenbrane] (L2) at (-2.9,0) {};
\node[sevenbrane] (L1) at (-1.75,0) {};

\node[sevenbrane] (R1) at (1.75,0) {};
\node[sevenbrane] (R2) at (2.9,0) {};
\node[sevenbrane] (R3) at (4.0,0) {};

\node[sevenbrane] (U3) at (0,4.0) {};
\node[sevenbrane] (U2) at (0,2.9) {};
\node[sevenbrane] (U1) at (0,1.75) {};

\node[sevenbrane] (D1) at (0,-1.75) {};
\node[sevenbrane] (D2) at (0,-2.9) {};
\node[sevenbrane] (D3) at (0,-4.0) {};



\draw[brane] (L3) -- (L2);

\draw[brane] (-2.9,0.07) -- (-1.75,0.07);
\draw[brane] (-2.9,-0.07) -- (-1.75,-0.07);

\draw[brane] (-1.75,0.16) -- (1.75,0.16);
\draw[brane] (-1.75,0) -- (1.75,0);
\draw[brane] (-1.75,-0.16) -- (1.75,-0.16);

\draw[brane] (1.75,0.07) -- (2.9,0.07);
\draw[brane] (1.75,-0.07) -- (2.9,-0.07);

\draw[brane] (R2) -- (R3);



\draw[brane] (U3) -- (U2);

\draw[brane] (0.07,2.9) -- (0.07,1.75);
\draw[brane] (-0.07,2.9) -- (-0.07,1.75);

\draw[brane] (0.16,1.75) -- (0.16,-1.75);
\draw[brane] (0,1.75) -- (0,-1.75);
\draw[brane] (-0.16,1.75) -- (-0.16,-1.75);

\draw[brane] (0.07,-1.75) -- (0.07,-2.9);
\draw[brane] (-0.07,-1.75) -- (-0.07,-2.9);

\draw[brane] (D2) -- (D3);

\foreach \p in {L3,L2,L1,R1,R2,R3,U3,U2,U1,D1,D2,D3}{
    \node[sevenbrane] at (\p) {};
}

\end{tikzpicture}
\caption{Infinite-coupling (p,q)-web for the $\mathcal T_{3,3}$ theory.}
\label{fig:su3-su3-infinite-coupling-web}
\end{figure}
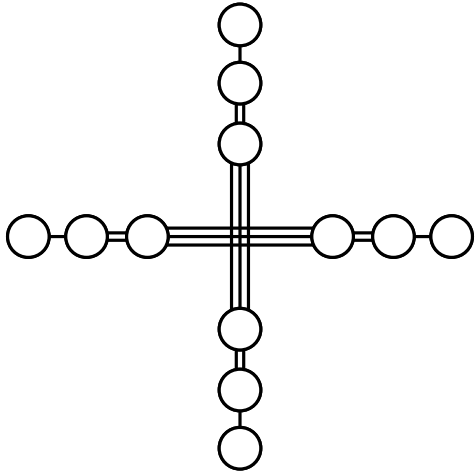

In this section we present an alternative derivation of the Hasse diagram for the Higgs branch of the $\mathcal{T}_{3,3}$ theory. In this case, both $\vQ{n}{k},   \hQ{n}{k}$ read
\begin{equation}
\begin{gathered}
\label{eq:electricquivert33}
  \begin{tikzpicture}
        \node[flavourSU,label=below:{$k$}] (0) at (0,0) {};
        \node[gaugeSU,label=below:{$k$}] (1) at (1,0) {};
        \node (2) at (2,0) {$\cdots$};
        \node[gaugeSU,label=below:{$k$}] (3) at (3,0) {};
        \node[flavourSU,label=below:{$k$}] (4) at (4,0) {};
        \draw (0)--(1)--(2)--(3)--(4);
        \draw [decorate,decoration={brace,amplitude=5pt},xshift=0pt,yshift=0pt] (3.1,-0.8)--(1-0.1,-0.8) node [black,midway,yshift=-0.5cm] {$n-1$ nodes};
    \end{tikzpicture}
    \end{gathered}
\end{equation}
The UV fixed point can also be described via the  (p,q)-web in Figure \ref{fig:su3-su3-infinite-coupling-web} 
By a standard procedure, we can read the magnetic quiver from the (p,q) web, obtaining the following result: 
\begin{equation}\label{fig:full UV T33}
    \mathsf{Q}_m^{\infty}=\quad\raisebox{-.5\height}{\scalebox{0.7}{\begin{tikzpicture}
                \node[gaugeBig,label=below:{$1$}] (1) at (0,0) {};
                \node[gaugeBig,label=below:{$2$}] (2) at (2,0) {};
                \node[gaugeBig,label=below:{$3$}] (3) at (4,0) {};
                \node[gaugeBig,label=below:{$2$}] (4) at (6,0) {};
                \node[gaugeBig,label=below:{$1$}] (5) at (8,0) {};
                \node[gaugeBig,label=above:{$1$}] (3f1) at (0,2) {};
                \node[gaugeBig,label=above:{$2$}] (3f2) at (2,1.5) {};
                \node[gaugeBig,label=above:{$3$}] (3f3) at (4,1) {};
                \node[gaugeBig,label=above:{$2$}] (3f4) at (6,1.5) {};
                \node[gaugeBig,label=above:{$1$}] (3f5) at (8,2) {};
                \draw (1)--(2)--(3)--(4)--(5) (3)--(3f3) (3f1)--(3f2)--(3f3)--(3f4)--(3f5);
            \end{tikzpicture}}}\quad.
\end{equation}
To study the Higgs Branch Hasse diagram, we can exploit the decay and fission technique \cite{Bourget:2023dkj}. Drawing on top of the diagram the magnetic quiver for the SCFT, and on each node (corresponding to an Hasse diagram leaf) the magnetic quiver of the massless theory on that leaf, the result is the following:  
\begin{landscape}
\begin{equation}
    \begin{tikzpicture}
        \node (a) at (0,0) {$\raisebox{-.5\height}{\scalebox{0.4}{\begin{tikzpicture}
                \node[gaugeBig,label=below:{$1$}] (1) at (0,0) {};
                \node[gaugeBig,label=below:{$2$}] (2) at (2,0) {};
                \node[gaugeBig,label=below:{$3$}] (3) at (4,0) {};
                \node[gaugeBig,label=below:{$2$}] (4) at (6,0) {};
                \node[gaugeBig,label=below:{$1$}] (5) at (8,0) {};
                \node[gaugeBig,label=above:{$1$}] (3f1) at (0,2) {};
                \node[gaugeBig,label=above:{$2$}] (3f2) at (2,1.5) {};
                \node[gaugeBig,label=above:{$3$}] (3f3) at (4,1) {};
                \node[gaugeBig,label=above:{$2$}] (3f4) at (6,1.5) {};
                \node[gaugeBig,label=above:{$1$}] (3f5) at (8,2) {};
                \draw (1)--(2)--(3)--(4)--(5) (3)--(3f3) (3f1)--(3f2)--(3f3)--(3f4)--(3f5);
            \end{tikzpicture}}}$};
        \node (b1) at (-10,-3) {$\raisebox{-.5\height}{\scalebox{0.4}{\begin{tikzpicture}
                \node[gaugeBig,label=below:{$1$}] (1) at (0,0) {};
                \node[gaugeBig,label=below:{$2$}] (2) at (2,0) {};
                \node[gaugeBig,label=below:{$3$}] (3) at (4,0) {};
                \node[gaugeBig,label=below:{$2$}] (4) at (6,0) {};
                \node[gaugeBig,label=below:{$1$}] (5) at (8,0) {};
                \node[gaugeBig,label=above:{$1$}] (3f2) at (2,1.5) {};
                \node[gaugeBig,label=above:{$3$}] (3f3) at (4,1) {};
                \node[gaugeBig,label=above:{$2$}] (3f4) at (6,1.5) {};
                \node[gaugeBig,label=above:{$1$}] (3f5) at (8,2) {};
                \draw (1)--(2)--(3)--(4)--(5) (3)--(3f3) (3f2)--(3f3)--(3f4)--(3f5);
            \end{tikzpicture}}}$};
        \node (b2) at (-5,-3) {$\raisebox{-.5\height}{\scalebox{0.4}{\begin{tikzpicture}
                \node[gaugeBig,label=below:{$1$}] (1) at (0,0) {};
                \node[gaugeBig,label=below:{$2$}] (2) at (2,0) {};
                \node[gaugeBig,label=below:{$3$}] (3) at (4,0) {};
                \node[gaugeBig,label=below:{$2$}] (4) at (6,0) {};
                \node[gaugeBig,label=below:{$1$}] (5) at (8,0) {};
                \node[gaugeBig,label=above:{$1$}] (3f1) at (0,2) {};
                \node[gaugeBig,label=above:{$2$}] (3f2) at (2,1.5) {};
                \node[gaugeBig,label=above:{$3$}] (3f3) at (4,1) {};
                \node[gaugeBig,label=above:{$1$}] (3f4) at (6,1.5) {};
                \draw (1)--(2)--(3)--(4)--(5) (3)--(3f3) (3f1)--(3f2)--(3f3)--(3f4);
            \end{tikzpicture}}}$};
        \node (c1) at (5,-3) {$\raisebox{-.5\height}{\scalebox{0.4}{\begin{tikzpicture}
                \node[gaugeBig,label=below:{$1$}] (2) at (2,0) {};
                \node[gaugeBig,label=below:{$3$}] (3) at (4,0) {};
                \node[gaugeBig,label=below:{$2$}] (4) at (6,0) {};
                \node[gaugeBig,label=below:{$1$}] (5) at (8,0) {};
                \node[gaugeBig,label=above:{$1$}] (3f1) at (0,2) {};
                \node[gaugeBig,label=above:{$2$}] (3f2) at (2,1.5) {};
                \node[gaugeBig,label=above:{$3$}] (3f3) at (4,1) {};
                \node[gaugeBig,label=above:{$2$}] (3f4) at (6,1.5) {};
                \node[gaugeBig,label=above:{$1$}] (3f5) at (8,2) {};
                \draw (2)--(3)--(4)--(5) (3)--(3f3) (3f1)--(3f2)--(3f3)--(3f4)--(3f5);
            \end{tikzpicture}}}$};
        \node (c2) at (10,-3) {$\raisebox{-.5\height}{\scalebox{0.4}{\begin{tikzpicture}
                \node[gaugeBig,label=below:{$1$}] (1) at (0,0) {};
                \node[gaugeBig,label=below:{$2$}] (2) at (2,0) {};
                \node[gaugeBig,label=below:{$3$}] (3) at (4,0) {};
                \node[gaugeBig,label=below:{$1$}] (4) at (6,0) {};
                \node[gaugeBig,label=above:{$1$}] (3f1) at (0,2) {};
                \node[gaugeBig,label=above:{$2$}] (3f2) at (2,1.5) {};
                \node[gaugeBig,label=above:{$3$}] (3f3) at (4,1) {};
                \node[gaugeBig,label=above:{$2$}] (3f4) at (6,1.5) {};
                \node[gaugeBig,label=above:{$1$}] (3f5) at (8,2) {};
                \draw (1)--(2)--(3)--(4) (3)--(3f3) (3f1)--(3f2)--(3f3)--(3f4)--(3f5);
            \end{tikzpicture}}}$};
        \node (d) at (-10,-6) {$\raisebox{-.5\height}{\scalebox{0.4}{\begin{tikzpicture}
                \node[gaugeBig,label=below:{$1$}] (1) at (0,0) {};
                \node[gaugeBig,label=below:{$2$}] (2) at (2,0) {};
                \node[gaugeBig,label=below:{$3$}] (3) at (4,0) {};
                \node[gaugeBig,label=below:{$2$}] (4) at (6,0) {};
                \node[gaugeBig,label=below:{$1$}] (5) at (8,0) {};
                \node[gaugeBig,label=above:{$1$}] (3f2) at (2,1.5) {};
                \node[gaugeBig,label=above:{$2$}] (3f3) at (4,1) {};
                \node[gaugeBig,label=above:{$1$}] (3f4) at (6,1.5) {};
                \draw (1)--(2)--(3)--(4)--(5) (3)--(3f3) (3f2)--(3f3)--(3f4);
            \end{tikzpicture}}}$};
        \node (e1) at (-5.5,-6) {$\raisebox{-.5\height}{\scalebox{0.4}{\begin{tikzpicture}
                \node[gaugeBig,label=below:{$1$}] (2) at (2,0) {};
                \node[gaugeBig,label=below:{$3$}] (3) at (4,0) {};
                \node[gaugeBig,label=below:{$2$}] (4) at (6,0) {};
                \node[gaugeBig,label=below:{$1$}] (5) at (8,0) {};
                \node[gaugeBig,label=above:{$1$}] (3f2) at (2,1.5) {};
                \node[gaugeBig,label=above:{$3$}] (3f3) at (4,1) {};
                \node[gaugeBig,label=above:{$2$}] (3f4) at (6,1.5) {};
                \node[gaugeBig,label=above:{$1$}] (3f5) at (8,2) {};
                \draw (2)--(3)--(4)--(5) (3)--(3f3) (3f2)--(3f3)--(3f4)--(3f5);
            \end{tikzpicture}}}$};
        \node (e2) at (-2,-6) {$\raisebox{-.5\height}{\scalebox{0.4}{\begin{tikzpicture}
                \node[gaugeBig,label=below:{$1$}] (2) at (2,0) {};
                \node[gaugeBig,label=below:{$3$}] (3) at (4,0) {};
                \node[gaugeBig,label=below:{$2$}] (4) at (6,0) {};
                \node[gaugeBig,label=below:{$1$}] (5) at (8,0) {};
                \node[gaugeBig,label=above:{$1$}] (3f1) at (0,2) {};
                \node[gaugeBig,label=above:{$2$}] (3f2) at (2,1.5) {};
                \node[gaugeBig,label=above:{$3$}] (3f3) at (4,1) {};
                \node[gaugeBig,label=above:{$1$}] (3f4) at (6,1.5) {};
                \draw (2)--(3)--(4)--(5) (3)--(3f3) (3f1)--(3f2)--(3f3)--(3f4);
            \end{tikzpicture}}}$};
        \node (e3) at (2,-6) {$\raisebox{-.5\height}{\scalebox{0.4}{\begin{tikzpicture}
                \node[gaugeBig,label=below:{$1$}] (1) at (0,0) {};
                \node[gaugeBig,label=below:{$2$}] (2) at (2,0) {};
                \node[gaugeBig,label=below:{$3$}] (3) at (4,0) {};
                \node[gaugeBig,label=below:{$1$}] (4) at (6,0) {};
                \node[gaugeBig,label=above:{$1$}] (3f2) at (2,1.5) {};
                \node[gaugeBig,label=above:{$3$}] (3f3) at (4,1) {};
                \node[gaugeBig,label=above:{$2$}] (3f4) at (6,1.5) {};
                \node[gaugeBig,label=above:{$1$}] (3f5) at (8,2) {};
                \draw (1)--(2)--(3)--(4) (3)--(3f3) (3f2)--(3f3)--(3f4)--(3f5);
            \end{tikzpicture}}}$};
        \node (e4) at (5.5,-6) {$\raisebox{-.5\height}{\scalebox{0.4}{\begin{tikzpicture}
                \node[gaugeBig,label=below:{$1$}] (1) at (0,0) {};
                \node[gaugeBig,label=below:{$2$}] (2) at (2,0) {};
                \node[gaugeBig,label=below:{$3$}] (3) at (4,0) {};
                \node[gaugeBig,label=below:{$1$}] (4) at (6,0) {};
                \node[gaugeBig,label=above:{$1$}] (3f1) at (0,2) {};
                \node[gaugeBig,label=above:{$2$}] (3f2) at (2,1.5) {};
                \node[gaugeBig,label=above:{$3$}] (3f3) at (4,1) {};
                \node[gaugeBig,label=above:{$1$}] (3f4) at (6,1.5) {};
                \draw (1)--(2)--(3)--(4) (3)--(3f3) (3f1)--(3f2)--(3f3)--(3f4);
            \end{tikzpicture}}}$};
        \node (f) at (10,-6) {$\raisebox{-.5\height}{\scalebox{0.4}{\begin{tikzpicture}
                \node[gaugeBig,label=below:{$1$}] (2) at (2,0) {};
                \node[gaugeBig,label=below:{$2$}] (3) at (4,0) {};
                \node[gaugeBig,label=below:{$1$}] (4) at (6,0) {};
                \node[gaugeBig,label=above:{$1$}] (3f1) at (0,2) {};
                \node[gaugeBig,label=above:{$2$}] (3f2) at (2,1.5) {};
                \node[gaugeBig,label=above:{$3$}] (3f3) at (4,1) {};
                \node[gaugeBig,label=above:{$2$}] (3f4) at (6,1.5) {};
                \node[gaugeBig,label=above:{$1$}] (3f5) at (8,2) {};
                \draw (2)--(3)--(4) (3)--(3f3) (3f1)--(3f2)--(3f3)--(3f4)--(3f5);
            \end{tikzpicture}}}$};
        \node (g1) at (-10,-9) {$\raisebox{-.5\height}{\scalebox{0.4}{\begin{tikzpicture}
                \node[gaugeBig,label=below:{$1$}] (1) at (0,0) {};
                \node[gaugeBig,label=below:{$2$}] (2) at (2,0) {};
                \node[gaugeBig,label=below:{$3$}] (3) at (4,0) {};
                \node[gaugeBig,label=below:{$2$}] (4) at (6,0) {};
                \node[gaugeBig,label=below:{$1$}] (5) at (8,0) {};
                \node[gaugeBig,label=above:{$2$}] (3f3) at (4,1) {};
                \node[gaugeBig,label=above:{$1$}] (3f4) at (6,1.5) {};
                \draw (1)--(2)--(3)--(4)--(5) (3)--(3f3) (3f3)--(3f4);
            \end{tikzpicture}}}$};
        \node (g2) at (-5,-9) {$\raisebox{-.5\height}{\scalebox{0.4}{\begin{tikzpicture}
                \node[gaugeBig,label=below:{$1$}] (1) at (0,0) {};
                \node[gaugeBig,label=below:{$2$}] (2) at (2,0) {};
                \node[gaugeBig,label=below:{$3$}] (3) at (4,0) {};
                \node[gaugeBig,label=below:{$2$}] (4) at (6,0) {};
                \node[gaugeBig,label=below:{$1$}] (5) at (8,0) {};
                \node[gaugeBig,label=above:{$1$}] (3f2) at (2,1.5) {};
                \node[gaugeBig,label=above:{$2$}] (3f3) at (4,1) {};
                \draw (1)--(2)--(3)--(4)--(5) (3)--(3f3) (3f2)--(3f3);
            \end{tikzpicture}}}$};
        \node (h) at (0,-9) {$\raisebox{-.5\height}{\scalebox{0.4}{\begin{tikzpicture}
                \node[gaugeBig,label=below:{$1$}] (2) at (2,0) {};
                \node[gaugeBig,label=below:{$2$}] (3) at (4,0) {};
                \node[gaugeBig,label=below:{$1$}] (4) at (6,0) {};
                \node[gaugeBig,label=above:{$1$}] (3f2) at (2,1.5) {};
                \node[gaugeBig,label=above:{$2$}] (3f3) at (4,1) {};
                \node[gaugeBig,label=above:{$1$}] (3f4) at (6,1.5) {};
                \draw (2)--(3)--(4) (3)--(3f3) (3f2)--(3f3)--(3f4);
            \end{tikzpicture}}}$};
        \node (i1) at (5,-9) {$\raisebox{-.5\height}{\scalebox{0.4}{\begin{tikzpicture}
                \node[gaugeBig,label=below:{$2$}] (3) at (4,0) {};
                \node[gaugeBig,label=below:{$1$}] (4) at (6,0) {};
                \node[gaugeBig,label=above:{$1$}] (3f1) at (0,2) {};
                \node[gaugeBig,label=above:{$2$}] (3f2) at (2,1.5) {};
                \node[gaugeBig,label=above:{$3$}] (3f3) at (4,1) {};
                \node[gaugeBig,label=above:{$2$}] (3f4) at (6,1.5) {};
                \node[gaugeBig,label=above:{$1$}] (3f5) at (8,2) {};
                \draw (3)--(4) (3)--(3f3) (3f1)--(3f2)--(3f3)--(3f4)--(3f5);
            \end{tikzpicture}}}$};
        \node (i2) at (10,-9) {$\raisebox{-.5\height}{\scalebox{0.4}{\begin{tikzpicture}
                \node[gaugeBig,label=below:{$1$}] (2) at (2,0) {};
                \node[gaugeBig,label=below:{$2$}] (3) at (4,0) {};
                \node[gaugeBig,label=above:{$1$}] (3f1) at (0,2) {};
                \node[gaugeBig,label=above:{$2$}] (3f2) at (2,1.5) {};
                \node[gaugeBig,label=above:{$3$}] (3f3) at (4,1) {};
                \node[gaugeBig,label=above:{$2$}] (3f4) at (6,1.5) {};
                \node[gaugeBig,label=above:{$1$}] (3f5) at (8,2) {};
                \draw (2)--(3) (3)--(3f3) (3f1)--(3f2)--(3f3)--(3f4)--(3f5);
            \end{tikzpicture}}}$};
        \node (j) at (0,-12) {$\raisebox{-.5\height}{\scalebox{0.4}{\begin{tikzpicture}
                \node[gaugeBig,label=below:{$1$}] (3) at (4,0) {};
            \end{tikzpicture}}}$};
        \draw[->] (a)--(b1);
        \draw[->] (a)--(b2);
        \draw[->] (a)--(c1);
        \draw[->] (a)--(c2);
        \draw[->] (b1)--(d);
        \draw[->] (b2)--(d);
        \draw[->] (c1)--(f);
        \draw[->] (c2)--(f);
        \draw[->] (b1)--(e1);
        \draw[->] (b1) .. controls (-5.5,-4.5) and (-1,-4.5) .. (e3);
        \draw[->] (b2)--(e2);
        \draw[->] (b2) .. controls (2,-4) .. (e4);
        \draw[->] (c1) .. controls (-1.5,-4) .. (e1);
        \draw[->] (c1) .. controls (0,-4) .. (e2);
        \draw[->] (c2) .. controls (7.5,-4.5) and (3,-4.5) .. (e3);
        \draw[->] (c2)--(e4);
        \draw[->] (e1)--(h);
        \draw[->] (e2)--(h);
        \draw[->] (e3)--(h);
        \draw[->] (e4)--(h);
        \draw[->] (d)--(g1);
        \draw[->] (d)--(g2);
        \draw[->] (d)--(h);
        \draw[->] (f)--(h);
        \draw[->] (f)--(i1);
        \draw[->] (f)--(i2);
        \draw[->] (g1)--(j);
        \draw[->] (g2)--(j);
        \draw[->] (h)--(j);
        \draw[->] (i1)--(j);
        \draw[->] (i2)--(j);
    \end{tikzpicture}
\end{equation}
\end{landscape}
The corresponding transitions in the Hasse diagrams are:
\begin{equation}\label{fig:Ex2QuivSubt}
    \begin{tikzpicture}
            \node[hasse] (a) at (0,0) {};
            \node[hasse] (cl1) at (-7,4) {};
            \node[hasse] (cl2) at (-5,4) {};
            \node[hasse] (cm) at (0,4) {};
            \node[hasse] (c1) at (5,4) {};
            \node[hasse] (c2) at (7,4) {};
            \node[hasse] (dl) at (-6,6) {};
            \node[hasse] (dm1) at (-3,6) {};
            \node[hasse] (dm2) at (-1,6) {};
            \node[hasse] (dm3) at (1,6) {};
            \node[hasse] (dm4) at (3,6) {};
            \node[hasse] (d) at (6,6) {};
            \node[hasse] (el1) at (-5,8) {};
            \node[hasse] (el2) at (-3,8) {};
            \node[hasse] (e1) at (3,8) {};
            \node[hasse] (e2) at (5,8) {};
            \node[hasse] (f) at (0,10) {};
            \draw[cyan] (a)--(c1)--(d)--(c2)--(a) (d)--(e1)--(f)--(e2)--(d);
            \node at ($(a)!0.5!(c1)$) {$e_6$};
            \node at ($(a)!0.5!(c2)$) {$e_6$};
            \node at ($(d)!0.5!(c1)$) {$a_1$};
            \node at ($(d)!0.5!(c2)$) {$a_1$};
            \node at ($(d)!0.5!(e1)$) {$a_3$};
            \node at ($(d)!0.5!(e2)$) {$a_3$};
            \node at ($(f)!0.5!(e1)$) {$a_2$};
            \node at ($(f)!0.5!(e2)$) {$a_2$};
            \draw[gray] (a)--(cm)--(dl) (cm)--(dm1) (cm)--(dm2) (cm)--(dm3) (cm)--(dm4) (cm)--(d);
            \node at ($(a)!0.5!(cm)$) {$d_5$};
            \node at ($(cm)!0.5!(dl)$) {$a_3$};
            \node at ($(cm)!0.5!(dm1)$) {$a_5$};
            \node at ($(cm)!0.5!(dm2)$) {$a_5$};
            \node at ($(cm)!0.5!(dm3)$) {$a_5$};
            \node at ($(cm)!0.5!(dm4)$) {$a_5$};
            \node at ($(cm)!0.5!(d)$) {$a_3$};
            \draw[red] (a)--(cl1)--(dl)--(cl2)--(a);
            \node at ($(a)!0.5!(cl1)$) {$e_6$};
            \node at ($(a)!0.5!(cl2)$) {$e_6$};
            \node at ($(dl)!0.5!(cl1)$) {$a_1$};
            \node at ($(dl)!0.5!(cl2)$) {$a_1$};
            \draw[red] (f)--(el1)--(dl)--(el2)--(f);
            \node at ($(dl)!0.5!(el1)$) {$a_3$};
            \node at ($(dl)!0.5!(el2)$) {$a_3$};
            \node at ($(f)!0.5!(el1)$) {$a_2$};
            \node at ($(f)!0.5!(el2)$) {$a_2$};
            \draw[gray] (dm1)--(el1)--(dm2);
            \node at ($(dm1)!0.5!(el1)$) {$a_2$};
            \node at ($(dm2)!0.5!(el1)$) {$a_2$};
            \draw[gray] (dm3)--(el2)--(dm4);
            \node at ($(dm3)!0.5!(el2)$) {$a_2$};
            \node at ($(dm4)!0.7!(el2)$) {$a_2$};
            \draw[gray] (dm1)--(e1) (e2)--(dm2);
            \node at ($(dm1)!0.7!(e1)$) {$a_2$};
            \node at ($(dm2)!0.1!(e2)$) {$a_2$};
            \draw[gray] (dm3)--(e1) (e2)--(dm4);
            \node at ($(dm3)!0.7!(e1)$) {$a_2$};
            \node at ($(dm4)!0.4!(e2)$) {$a_2$};
        \end{tikzpicture},
\end{equation}
where we used the same color code of \eqref{Hasse complete T33}. 
Each leaf of this Hasse diagram clearly corresponds to a geometric deformation of the CY3 by matching this diagram with our geometric Hasse diagram in \eqref{Hasse complete T33}. Moreover, each dynamical deformation matches with a movement of a 5-brane in the corresponding web, as depicted in Figure \ref{ref: box brane web def}.

\begin{figure}[H]
\centering
\includegraphics[width=12cm]{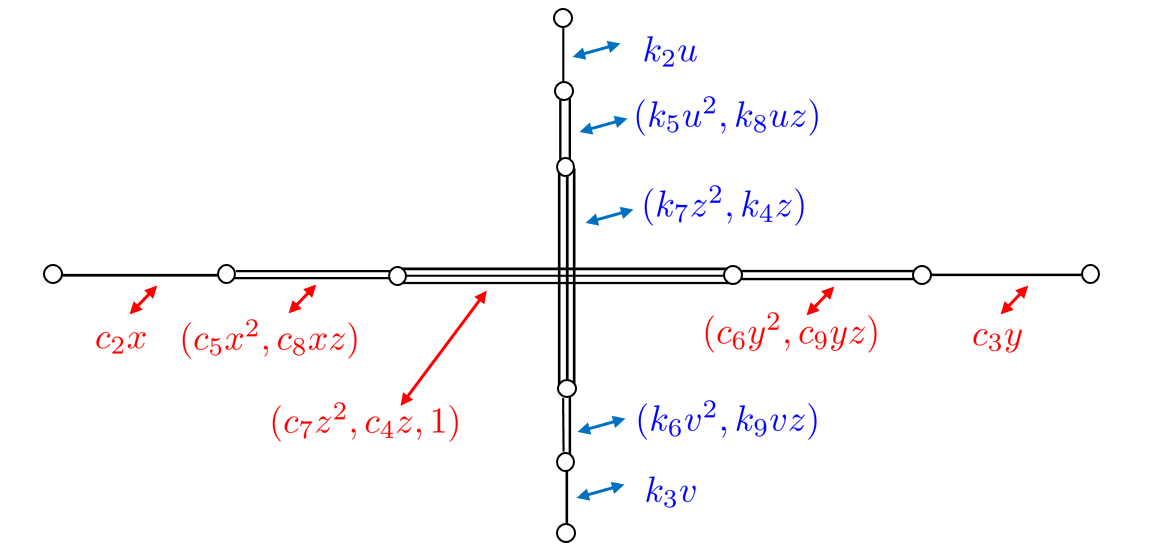}
\caption{Correspondence between branes movements and dynamical deformations in the $\mathcal{T}_{3,3}$ case.}
\label{ref: box brane web def}
\end{figure}

\section{Another detailed example: the case of the $T_4$ theory}\label{sec: Tn theories}
In this section, we will study another example of our geometric construction: the 5d $T_n$ theories. 
\subsection{$T_n$ theories and geometry}
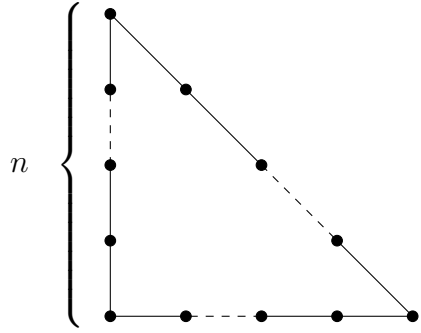
\begin{figure}[H]
\centering
    \scalebox{1.}{
    \begin{tikzpicture}
        \node at (-1.2,2) {$n$};
        \node at (-0.3,2) {$\begin{cases}
            \\
            \\
            \\
            \\
            \\
            \\
        \end{cases}$};
        \draw (0,4)--(1,3)--(2,2);
        \draw[dashed] (2,2)--(3,1);
        \draw (3,1)--(4,0)--(2,0);
        \draw[dashed] (2,0)--(1,0);
        \draw (1,0)--(0,0)--(0,2);
        \draw[dashed] (0,2)--(0,3);
        \draw (0,3)--(0,4);
        \filldraw (0,0) circle (2pt);
        \filldraw (0,1) circle (2pt);
        \filldraw (1,0) circle (2pt);
        \filldraw (0,2) circle (2pt);
        \filldraw (0,3) circle (2pt);
        \filldraw (0,4) circle (2pt);
        \filldraw (1,3) circle (2pt);
        \filldraw (2,0) circle (2pt);
        \filldraw (2,2) circle (2pt);
        \filldraw (3,0) circle (2pt);
        \filldraw (3,1) circle (2pt);
        \filldraw (4,0) circle (2pt);
    \end{tikzpicture}}
    \caption{Toric diagram of the $T_n$ theories. The three sides have equal lattice length.}
    \label{fig:Tn}
    \end{figure}

The 5d SCFTs known as $T_n$ theories were first introduced via a brane-web construction in \cite{Benini:2009gi}. We represent the dual toric diagram in Figure \ref{fig:Tn}.
Alternatively, they can be engineered via M-theory on the orbifold $\mathbb{C}^3/(\mathbb{Z}_n\times\mathbb{Z}_n)$, which can be nicely displayed as a hypersurface equation:
\begin{equation}\label{Tn hypersurface}
    xyz = w^n \quad \subset \mathbb{C}^4.
\end{equation}
In order to inspect the deformation theory of \eqref{Tn hypersurface}, it is convenient to rewrite it as a $\mathbb{C}^*$-fibration. This is equivalent to choosing a Type IIA reduction of the M-theory configuration. E.g.\ one can pick:
\begin{equation}\label{Tn C star}
\begin{cases}
    x u = w^n\\
    yz = u\\
    \end{cases}, \quad \mathbb{C}^*-\text{action}: (x,y,z,u,w) \rightarrow (x,\lambda y,\lambda^{-1}z,u,w).
\end{equation}
One can proceed as in the previous section: 
\begin{itemize}
    \item perform a resolution of the first equation in \eqref{Tn C star}, which is of Du Val type $A_{n-1}$;
    \item Identify the brane locus, where the $\mathbb{C}^*$-fiber degenerates. In this case it reads:
    \begin{equation}\label{Tn brane locus}
        \Delta_{D6} = u|_{u=u(z_i,e_i)} = z_1^ne_1^{n-1}\ldots e_{n-1}, 
    \end{equation}
\end{itemize}
where we have rewritten the brane locus in terms of the blow-up coordinates. From the Type IIA description \eqref{Tn brane locus} one can hence read off a possible low-energy electric quiver phase for the $T_n$ theory:
\begin{equation}\label{Tn quiver}
        \scalebox{0.65}{\begin{tikzpicture}
    \draw[thick] (-0.7, -0.7) rectangle (0.7, 0.7);
    \node at (0, 0) {\small$n$};
    
    \draw[thick] (0.7, 0)--(1.3, 0);
    
    \draw[thick] (2, 0) circle (0.65);
    \node at (2, 0) {\small$n-1$};
    
    \draw[thick] (2.7, 0)--(3.3, 0);
    
    \draw[thick] (4, 0) circle (0.65);
    \node at (4, 0) {\small$n-2$};
    
    \draw[thick] (4.7, 0)--(5.3, 0);

    \draw[dashed] (5.3, 0)--(5.9, 0);

    \draw[thick] (6, 0)--(6.6, 0);
    
    \draw[thick] (7.3,0) circle (0.65);
    \node at (7.3, 0) {\small$1$};
\end{tikzpicture}}
\end{equation}
Computing the number of Higgs branch moduli from the low-energy quiver, and adding the obstructed modes that become massless once all the gauge couplings in \eqref{Tn quiver} are turned to infinity, we expect that the Higgs branch dimension of the $T_n$ UV fixed point is, recalling expression \eqref{UV HB} 
\begin{equation}\label{Tn HB dim}
    \text{dim}_{\mathbb{H}}(HB) = n_H -n_V - \text{rank}(A_{n-1}) +\text{dim}(A_{n-1}) = \frac{(n-1)(3n+2)}{2}.
\end{equation}
This is in agreement with the analysis stemming from the magnetic quiver at infinite coupling, that reads:
\begin{equation}
    \raisebox{-.5\height}{\scalebox{0.7}{\begin{tikzpicture}
                \node[gaugeBig,label=below:{$1$}] (1) at (0,0) {};
                \node[gaugeBig,label=below:{$2$}] (2) at (2,0) {};
                \node[gaugeBig,label=below:{$n$}] (3) at (4,0) {};
                \node[gaugeBig,label=below:{$2$}] (4) at (6,0) {};
                \node[gaugeBig,label=below:{$1$}] (5) at (8,0) {};
                \node[gaugeBig,label=left:{$2$}] (6) at (4,2) {};
                \node[gaugeBig,label=left:{$1$}] (7) at (4,4) {};
                \draw (1)--(2);
                \draw[dashed] (2)--(3)--(4);
                \draw (4)--(5);
                \draw[dashed] (3)--(6);
                \draw (6)--(7);
            \end{tikzpicture}}}\;.
\end{equation}
In order to make contact with the algebraic point of view, we can compute the normalizable deformations of the second equation in \eqref{Tn C star}, exploiting the brane locus description. Uplifting the result back to the M-theory singular phase, we find that the normalizable deformations are:
\begin{equation}\label{Tn def}
    \begin{cases}
    x u = w^n\\
   yz = u + \sum\limits_{\substack{i,j \ge 0 \\ i+j \le n-2}}c_{ij}x^iw^j\\
    \end{cases}.
\end{equation}
Naturally, given the permutation symmetry $x\leftrightarrow y \leftrightarrow z$ in the defining equation of the $T_n$ theory \eqref{Tn hypersurface}, which is manifest from the toric diagram in Figure \ref{fig:Tn}, we could have also chosen two more inequivalent $\mathbb{C}^*$-fibrations:
\begin{equation}
\label{eq:otherreductionTn}
    \begin{cases}
        uz = w^n\\
        xy = u\\ 
    \end{cases}, \quad \begin{cases}
        uy = w^n\\
        xz = u\\ 
    \end{cases}.
\end{equation}
These give rise to further dynamical deformations of the form \eqref{Tn def}. Putting all the three inequivalent Type IIA descriptions together, and uplifting back to M-theory, we obtain that the deformed $T_n$ hypersurface equation is:
\begin{equation}\label{Tn hyp def}
    xyz = w^n + \sum\limits_{\substack{i,j \ge 0 \\ i+j \le n-2}}c_{ij}x^{i+1}w^j + \sum\limits_{\substack{i,j \ge 0 \\ i+j \le n-2}}\hat{c}_{ij}y^{i+1}w^j  + \sum\limits_{\substack{i,j \ge 0 \\ i+j \le n-2}}\tilde{c}_{ij}z^{i+1}w^j
\end{equation}
This is not quite the end of the story, though: \textit{we are missing $n$ dynamical deformations}, in order to match the expectation \eqref{Tn HB dim} on the dimension of the Higgs branch derived from the magnetic quiver and low-energy electric quiver perspective. \textit{These deformations cannot be detected via a Type IIA reduction, but are intrinsically M-theoretic.} Employing the permutation symmetry of the three singular lines of the $T_n$ theory, together with the data from the magnetic quiver, it is easy to argue that the missing deformations modify \eqref{Tn hyp def} as (appropriately redefining the coordinate $w$ in order to eliminate spurious deformations):
\begin{equation}\label{Tn hyp fully def}
    xyz = w^n + \sum\limits_{\substack{i,j \ge 0 \\ i+j \le n-2}}c_{ij}x^{i+1}w^j + \sum\limits_{\substack{i,j \ge 0 \\ i+j \le n-2}}\hat{c}_{ij}y^{i+1}w^j  + \sum\limits_{\substack{i,j \ge 0 \\ i+j \le n-2}}\tilde{c}_{ij}z^{i+1}w^j + \sum\limits_{i=1}^{n-1}\overline{c}_i w^i,
\end{equation}
where the intrinsically M-theoretic deformations appear in the last summand. 
From the point of view of the starting $T_n$ toric diagram, the deformations in the last summand of \eqref{Tn hyp fully def} trigger a RG flow to leaves that are effectively described by a toric diagram which has exactly the same shape as the $T_n$ one, but with shorter external edges. These leaves are nothing but the $T_{n-1}$,$T_{n-2},\ldots$ theories. In total, the number of dynamical deformations is:
\begin{equation}
    \# \text{ of dynamical deformations}(T_n) = \frac{(3n+2)(n-1)}{2}.
\end{equation}\\
\indent With the explicit expression of the deformed $T_n$ equation in our pockets, we can swiftly analyse the stratification of the Higgs branch, as we have previously shown for the theories in Section \ref{sec:UV HB box}. For the sake of concreteness, in the remaining part of this Section we will first explicitly compute the Hasse diagram of the $T_4$ theory from geometry, using the method explained in \cref{sec:stratificationfromgeometry},
and then we will check our result with the magnetic quiver.

\subsection{Geometric Hasse diagram of the $\mathcal T_4$ theory}
\begin{figure}[H]
    \centering
    \scalebox{0.7}{
    \begin{tikzpicture}
        \node (a) at (0,0) {$\emptyset$};

        \node[text=red]   (b) at (-8,-4)  {$(x^3,x^2w,x w^2)$};
        \node[text=green!60!black] (c) at (0,-4)   {$(y^3,y^2w,y w^2)$};
        \node[text=blue]  (d) at (8,-4)   {$(z^3,z^2w,z w^2)$};

        \node[text=red]   (e) at (-10,-9) {$(x^2,x w)$};
        \node             (f) at (-4,-9)  {$(w^3)$};
        \node[text=green!60!black] (g) at (4,-9)   {$(y^2,y w)$};
        \node[text=blue]  (h) at (10,-9)  {$(z^2,z w)$};

        \node (j) at (0,-13) {$1$};

        \draw[->,red] (a)--(b);
        \draw[->,green!60!black] (a)--(c);
        \draw[->,blue] (a)--(d);

        \draw[->, red] (b)--(e);
        \draw[->, green!60!black] (c)--(g);
        \draw[->, blue] (d)--(h);

        \draw[->] (b)--(f);
        \draw[->] (c)--(f);
        \draw[->] (d)--(f);

        \draw[->,red] (e)--(j);
        \draw[->] (f)--(j);
        \draw[->,green!60!black] (g)--(j);
        \draw[->,blue] (h)--(j);
    \end{tikzpicture}}
    \caption{Geometric Hasse diagram of the $T_4$ theory, with leaves labeled by complex deformations.}
    \label{fig:T4 hasse deformations}
\end{figure}
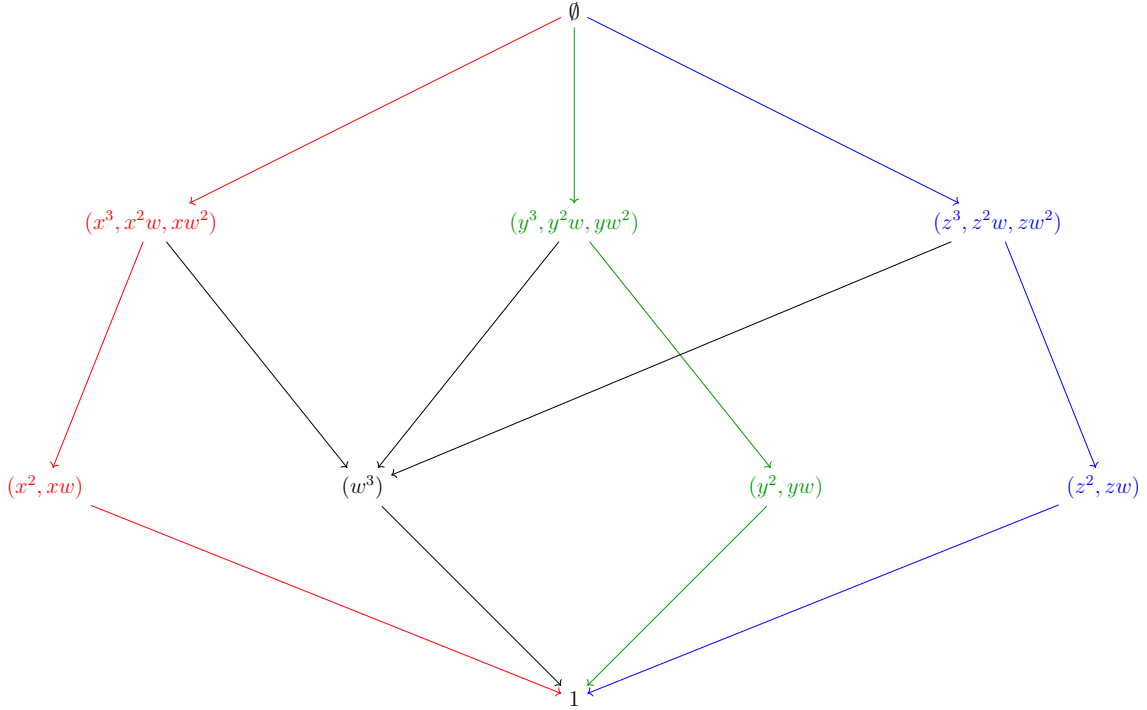

The singularity is obtained considering the $n = 4$ case in \eqref{Tn hypersurface}:
\begin{equation}
\label{eq:t4equation}
    x y z = w^4 \quad \subset \mathbb C^4,
\end{equation}
First, let us list the three curves of non-isolated $A_3$ singularities associated to the three edges of the triangle \Cref{fig:Tn}. These curves are located, respectively, at $x = y = w = 0,$ $x = z = w = 0,$ and $y = z = w = 0$. 
Given one of such curves, we can always go to IIA by reducing along a direction parallel  to the corresponding edge of the toric diagram. This corresponds, for the $y = z = w = 0$,  to the $\mathbb C^*$ fibration described in \eqref{Tn C star}, and for $x = y = w = 0,$ $x = z = w = 0,$ to the fibrations described in \eqref{eq:otherreductionTn}.  Let us start by the first curve $y = z = w = 0$. By exploiting the methods  outlined in  \cref{sec: HB quivers T33} and below equation \eqref{Tn C star}, we organize the deformations of the second equation of \eqref{Tn C star} as follows: 
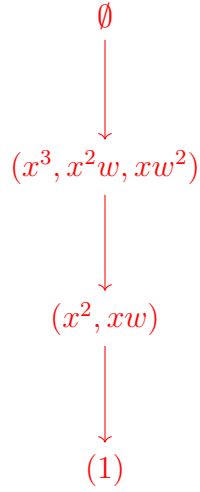
\begin{figure}[H]
    \centering
    \begin{tikzpicture}[
        every node/.style={text=red},
        every path/.style={draw=red, ->}
    ]
        \node (a) at (0,0) {$\emptyset$};
        \node (b) at (0,-2) {$(x^3,x^2w,x w^2)$};
        \node (c) at (0,-4) {$(x^2,x w)$};
        \node (d) at (0,-6) {$(1)$};

        \draw (a)--(b);
        \draw (b)--(c);
        \draw (c)--(d);
    \end{tikzpicture}
    \caption{Hasse diagram of the $T_4$ deformations that deform the curve $y = z = w = 0$.}
    \label{fig:T4_hasse_deformations}
\end{figure}
We can proceed analogously for the other two curves, obtaining the following partial Hasse diagrams 
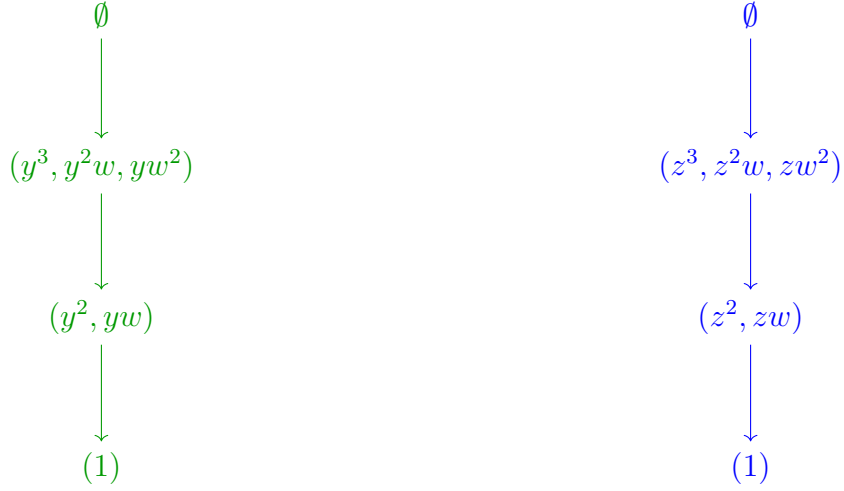
\begin{figure}[H]
    \centering

    \begin{minipage}{0.45\textwidth}
        \centering
        \begin{tikzpicture}[
            every node/.style={text=green!60!black},
            every path/.style={draw=green!60!black, ->}
        ]
            \node (a) at (0,0) {$\emptyset$};
            \node (b) at (0,-2) {$(y^3,y^2w,y w^2)$};
            \node (c) at (0,-4) {$(y^2,y w)$};
            \node (d) at (0,-6) {$(1)$};

            \draw (a)--(b);
            \draw (b)--(c);
            \draw (c)--(d);
        \end{tikzpicture}
    \end{minipage}
    \hfill
    \begin{minipage}{0.45\textwidth}
        \centering
        \begin{tikzpicture}[
            every node/.style={text=blue},
            every path/.style={draw=blue, ->}
        ]
            \node (a) at (0,0) {$\emptyset$};
            \node (b) at (0,-2) {$(z^3,z^2w,z w^2)$};
            \node (c) at (0,-4) {$(z^2,z w)$};
            \node (d) at (0,-6) {$(1)$};

            \draw (a)--(b);
            \draw (b)--(c);
            \draw (c)--(d);
        \end{tikzpicture}
    \end{minipage}

    \caption{Partial Hasse diagrams associated with the curves $x = z = w = 0$ (in green) and $x = y = w = 0$ (in blue).}
    \label{fig:T4_hasse_two_copies}
\end{figure}
To extract the HB of the 5d SCFT, we need to combine, as outlined in \cref{sec:stratificationfromgeometry}, \Cref{fig:T4_hasse_deformations} and \Cref{fig:T4_hasse_two_copies}. 
The resulting geometric Hasse diagram for the $T_4$ theory is depicted in \Cref{fig:T4 hasse deformations}, 

 We label each leaf in Figure \ref{fig:T4 hasse deformations} with the monomials that have to be turned on in \eqref{Tn hyp fully def} in order to land on it. Furthermore, black color was used for the node associated with the deformation $w^3$, which is one of the purely M-theoretic deformations described around \eqref{Tn hyp fully def}. Hasse diagrams for different values of $n$ can be computed analogously.

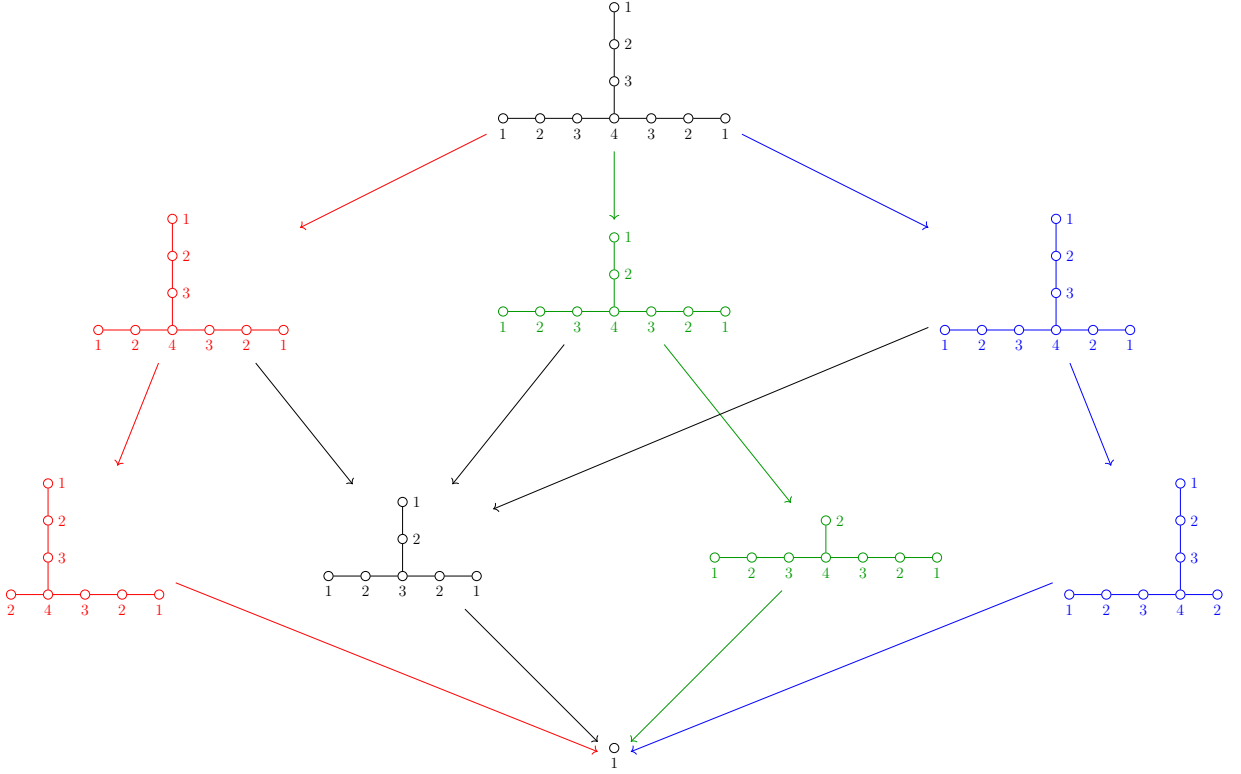
\begin{figure}[H]
    \centering
    \scalebox{0.7}{\begin{tikzpicture}
        \node (a) at (0,0) {$\raisebox{-.5\height}{\scalebox{0.7}{\begin{tikzpicture}
            \node[gauge,label=below:{$4$}] (c1) at (0,0) {};
            \node[gauge,label=below:{$3$}] (t1) at (1,0) {};
            \node[gauge,label=below:{$2$}] (t2) at (2,0) {};
            \node[gauge,label=below:{$1$}] (t3) at (3,0) {};
            \node[gauge,label=below:{$3$}] (q1) at (-1,0) {};
            \node[gauge,label=below:{$2$}] (q2) at (-2,0) {};
            \node[gauge,label=below:{$1$}] (q3) at (-3,0) {};
            \node[gauge,label=right:{$3$}] (r1) at (0,1) {};
            \node[gauge,label=right:{$2$}] (r2) at (0,2) {};
            \node[gauge,label=right:{$1$}] (r3) at (0,3) {};
            \draw (c1)--(t1)--(t2)--(t3) (c1)--(q1)--(q2)--(q3) (c1)--(r1)--(r2)--(r3);
        \end{tikzpicture}}}$};

        \node (b) at (-8,-4) {$\raisebox{-.5\height}{\scalebox{0.7}{\begin{tikzpicture}[red]
            \node[gauge,label=below:{$4$}] (c1) at (0,0) {};
            \node[gauge,label=below:{$3$}] (t1) at (1,0) {};
            \node[gauge,label=below:{$2$}] (t2) at (2,0) {};
            \node[gauge,label=below:{$1$}] (t3) at (3,0) {};
            \node[gauge,label=below:{$2$}] (q1) at (-1,0) {};
            \node[gauge,label=below:{$1$}] (q2) at (-2,0) {};
            \node[gauge,label=right:{$3$}] (r1) at (0,1) {};
            \node[gauge,label=right:{$2$}] (r2) at (0,2) {};
            \node[gauge,label=right:{$1$}] (r3) at (0,3) {};
            \draw (c1)--(t1)--(t2)--(t3) (c1)--(q1)--(q2) (c1)--(r1)--(r2)--(r3);
        \end{tikzpicture}}}$};

        \node (c) at (0,-4) {$\raisebox{-.5\height}{\scalebox{0.7}{\begin{tikzpicture}[green!60!black]
            \node[gauge,label=below:{$4$}] (c1) at (0,0) {};
            \node[gauge,label=below:{$3$}] (t1) at (1,0) {};
            \node[gauge,label=below:{$2$}] (t2) at (2,0) {};
            \node[gauge,label=below:{$1$}] (t3) at (3,0) {};
            \node[gauge,label=below:{$3$}] (q1) at (-1,0) {};
            \node[gauge,label=below:{$2$}] (q2) at (-2,0) {};
            \node[gauge,label=below:{$1$}] (q3) at (-3,0) {};
            \node[gauge,label=right:{$2$}] (r1) at (0,1) {};
            \node[gauge,label=right:{$1$}] (r2) at (0,2) {};
            \draw (c1)--(t1)--(t2)--(t3) (c1)--(q1)--(q2)--(q3) (c1)--(r1)--(r2);
        \end{tikzpicture}}}$};

        \node (d) at (8,-4) {$\raisebox{-.5\height}{\scalebox{0.7}{\begin{tikzpicture}[blue]
            \node[gauge,label=below:{$4$}] (c1) at (0,0) {};
            \node[gauge,label=below:{$2$}] (t1) at (1,0) {};
            \node[gauge,label=below:{$1$}] (t2) at (2,0) {};
            \node[gauge,label=below:{$3$}] (q1) at (-1,0) {};
            \node[gauge,label=below:{$2$}] (q2) at (-2,0) {};
            \node[gauge,label=below:{$1$}] (q3) at (-3,0) {};
            \node[gauge,label=right:{$3$}] (r1) at (0,1) {};
            \node[gauge,label=right:{$2$}] (r2) at (0,2) {};
            \node[gauge,label=right:{$1$}] (r3) at (0,3) {};
            \draw (c1)--(t1)--(t2) (c1)--(q1)--(q2)--(q3) (c1)--(r1)--(r2)--(r3);
        \end{tikzpicture}}}$};

        \node (e) at (-10,-9) {$\raisebox{-.5\height}{\scalebox{0.7}{\begin{tikzpicture}[red]
            \node[gauge,label=below:{$4$}] (c1) at (0,0) {};
            \node[gauge,label=below:{$3$}] (t1) at (1,0) {};
            \node[gauge,label=below:{$2$}] (t2) at (2,0) {};
            \node[gauge,label=below:{$1$}] (t3) at (3,0) {};
            \node[gauge,label=below:{$2$}] (q1) at (-1,0) {};
            \node[gauge,label=right:{$3$}] (r1) at (0,1) {};
            \node[gauge,label=right:{$2$}] (r2) at (0,2) {};
            \node[gauge,label=right:{$1$}] (r3) at (0,3) {};
            \draw (c1)--(t1)--(t2)--(t3) (c1)--(q1) (c1)--(r1)--(r2)--(r3);
        \end{tikzpicture}}}$};

        \node (f) at (-4,-9) {$\raisebox{-.5\height}{\scalebox{0.7}{\begin{tikzpicture}
            \node[gauge,label=below:{$3$}] (c1) at (0,0) {};
            \node[gauge,label=below:{$2$}] (t1) at (1,0) {};
            \node[gauge,label=below:{$1$}] (t2) at (2,0) {};
            \node[gauge,label=below:{$2$}] (q1) at (-1,0) {};
            \node[gauge,label=below:{$1$}] (q2) at (-2,0) {};
            \node[gauge,label=right:{$2$}] (r1) at (0,1) {};
            \node[gauge,label=right:{$1$}] (r2) at (0,2) {};
            \draw (c1)--(t1)--(t2) (c1)--(q1)--(q2) (c1)--(r1)--(r2);
        \end{tikzpicture}}}$};

        \node (g) at (4,-9) {$\raisebox{-.5\height}{\scalebox{0.7}{\begin{tikzpicture}[green!60!black]
            \node[gauge,label=below:{$4$}] (c1) at (0,0) {};
            \node[gauge,label=below:{$3$}] (t1) at (1,0) {};
            \node[gauge,label=below:{$2$}] (t2) at (2,0) {};
            \node[gauge,label=below:{$1$}] (t3) at (3,0) {};
            \node[gauge,label=below:{$3$}] (q1) at (-1,0) {};
            \node[gauge,label=below:{$2$}] (q2) at (-2,0) {};
            \node[gauge,label=below:{$1$}] (q3) at (-3,0) {};
            \node[gauge,label=right:{$2$}] (r1) at (0,1) {};
            \draw (c1)--(t1)--(t2)--(t3) (c1)--(q1)--(q2)--(q3) (c1)--(r1);
        \end{tikzpicture}}}$};

        \node (h) at (10,-9) {$\raisebox{-.5\height}{\scalebox{0.7}{\begin{tikzpicture}[blue]
            \node[gauge,label=below:{$4$}] (c1) at (0,0) {};
            \node[gauge,label=below:{$2$}] (t1) at (1,0) {};
            \node[gauge,label=below:{$3$}] (q1) at (-1,0) {};
            \node[gauge,label=below:{$2$}] (q2) at (-2,0) {};
            \node[gauge,label=below:{$1$}] (q3) at (-3,0) {};
            \node[gauge,label=right:{$3$}] (r1) at (0,1) {};
            \node[gauge,label=right:{$2$}] (r2) at (0,2) {};
            \node[gauge,label=right:{$1$}] (r3) at (0,3) {};
            \draw (c1)--(t1) (c1)--(q1)--(q2)--(q3) (c1)--(r1)--(r2)--(r3);
        \end{tikzpicture}}}$};

        \node (j) at (0,-13) {$\raisebox{-.5\height}{\scalebox{0.7}{\begin{tikzpicture}
            \node[gauge,label=below:{$1$}] (c1) at (0,0) {};
        \end{tikzpicture}}}$};

        \draw[->,red] (a)--(b);
        \draw[->,green!60!black] (a)--(c);
        \draw[->,blue] (a)--(d);

        \draw[->,red] (b)--(e);
        \draw[->] (b)--(f);
        \draw[->] (c)--(f);
        \draw[->,green!60!black] (c)--(g);
        \draw[->] (d)--(f);
        \draw[->,blue] (d)--(h);

        \draw[->,red] (e)--(j);
        \draw[->] (f)--(j);
        \draw[->,green!60!black] (g)--(j);
        \draw[->,blue] (h)--(j);
    \end{tikzpicture}}\;
    \caption{Hasse diagram of the $T_4$ theory in terms of leaves labeled by magnetic quivers.}
    \label{fig:T4 hasse MQ}
\end{figure}

To conclude, we can check that our result reproduces the Hasse diagram that can be predicted by the magnetic quiver. We depicted the result in  \Cref{fig:T4 hasse MQ}, using the same color code of \Cref{fig:T4 hasse deformations}

\begin{figure}[H]
\centering
\includegraphics[width=12cm]{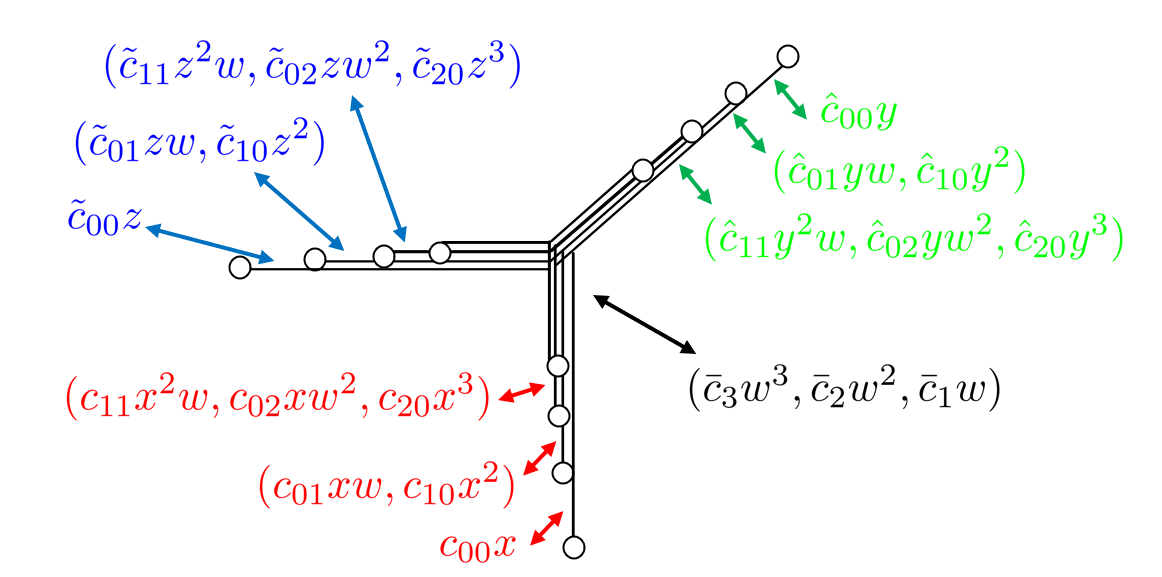}
\caption{Correspondence between branes movements and dynamical deformations in the $T_4$ case.}
\label{ref: Tn brane web def}
\end{figure}

We conclude by remarking that all the dynamical deformation monomials can be matched to movement of 5-branes in the corresponding $(p,q)$-web. For the $T_4$ case the match is as in Figure \ref{ref: Tn brane web def}.

\section{A comment on dynamical deformations and GTPs}\label{sec:white dots}
\begin{figure}[H]
\begin{center}
\scalemath{0.7}{
    \begin{tikzpicture}
         \node[bd] (00) at (-2.5,0) {};
        \node[bd] (10) at (-1.5,0) {};
        \node[bd] (20) at (-0.5,0) {};
        \node[bd] (30) at (0.5,0) {};
        \node[bd] (01) at (-2.5,1) {};
        \node[bd] (02) at (-2.5,2) {};
        \node[bd] (03) at (-2.5,3) {};
        \node[bd] (13) at (-1.5,3) {};
        \node[bd] (23) at (-0.5,3) {};
        \node[bd] (33) at (0.5,3) {};
        \node[bd] (31) at (0.5,1) {};
        \node[bd] (32) at (0.5,2) {};
        \draw (00)--(10)--(20)--(30)--(33)--(03)--(02)--(01)--(00);
        \node at (1.5,1.5) {$\longleftrightarrow$};
        \scalemath{0.8}{
        \draw[thick] (5,1.5)--(7,1.5) (5,1.6)--(7,1.6) (5,1.4)--(7,1.4);
        \draw[thick] (7,1.45)--(8,1.45) (7,1.55)--(8,1.55);
        \draw[thick] (4,1.45)--(5,1.45) (4,1.55)--(5,1.55);
        \draw[thick] (8,1.5)--(9,1.5);
        \draw[thick] (3,1.5)--(4,1.5);
        \draw[thick] (6,0.5)--(6,2.5) (5.9,0.5)--(5.9,2.5) (6.1,0.5)--(6.1,2.5);
        \draw[thick] (5.95,0.5)--(5.95,-0.5) (6.05,0.5)--(6.05,-0.5);
        \draw[thick] (5.95,2.5)--(5.95,3.5) (6.05,2.5)--(6.05,3.5);
        \draw[thick] (6,3.5)--(6,4.5);
        \draw[thick] (6,-0.5)--(6,-1.5);
        \node[bd, fill=white, draw=black] at (5,1.5) {};
        \node[bd, fill=white, draw=black] at (7,1.5) {};
        \node[bd, fill=white, draw=black] at (4,1.5) {};
        \node[bd, fill=white, draw=black] at (8,1.5) {};
        \node[bd, fill=white, draw=black] at (3,1.5) {};
        \node[bd, fill=white, draw=black] at (9,1.5) {};
        \node[bd, fill=white, draw=black] at (6,-1.5) {};
        \node[bd, fill=white, draw=black] at (6,-0.5) {};
        \node[bd, fill=white, draw=black] at (6,0.5) {};
        \node[bd, fill=white, draw=black] at (6,2.5) {};
        \node[bd, fill=white, draw=black] at (6,3.5) {};
        \node[bd, fill=white, draw=black] at (6,4.5) {};
        }
        \node[bd] (50) at (11,0) {};
        \node[bd] (60) at (12,0) {};
        \node[bd] (70) at (13,0) {};
        \node[bd] (80) at (14,0) {};
        \node[bd] (51) at (11,1) {};
        \node[bd, fill=white, draw=black] (52) at (11,2) {};
        \node[bd] (53) at (11,3) {};
        \node[bd] (63) at (12,3) {};
        \node[bd] (73) at (13,3) {};
        \node[bd] (83) at (14,3) {};
        \node[bd] (81) at (14,1) {};
        \node[bd] (82) at (14,2) {};
        \draw (50)--(60)--(70)--(80)--(83)--(53)--(52)--(51)--(50);
        \node at (15,1.5) {$\longleftrightarrow$};
         \scalemath{0.8}{
        \draw[thick] (21,1.5)--(23,1.5) (21,1.6)--(23,1.6) (21,1.4)--(23,1.4);
        \draw[thick] (23,1.45)--(24,1.45) (23,1.55)--(24,1.55);
        \draw[thick] (20,1.45)--(21,1.45) (20,1.55)--(21,1.55);
        \draw[thick] (24,1.5)--(25,1.5);
        \draw[thick] (22,0.5)--(22,2.5) (21.9,0.5)--(21.9,2.5) (22.1,0.5)--(22.1,2.5);
        \draw[thick] (21.95,0.5)--(21.95,-0.5) (22.05,0.5)--(22.05,-0.5);
        \draw[thick] (21.95,2.5)--(21.95,3.5) (22.05,2.5)--(22.05,3.5);
        \draw[thick] (22,3.5)--(22,4.5);
        \draw[thick] (22,-0.5)--(22,-1.5);
        \node[bd, fill=white, draw=black] at (21,1.5) {};
        \node[bd, fill=white, draw=black] at (23,1.5) {};
        \node[bd, fill=white, draw=black] at (20,1.5) {};
        \node[bd, fill=white, draw=black] at (24,1.5) {};
        \node[bd, fill=white, draw=black] at (25,1.5) {};
        \node[bd, fill=white, draw=black] at (22,-1.5) {};
        \node[bd, fill=white, draw=black] at (22,-0.5) {};
        \node[bd, fill=white, draw=black] at (22,0.5) {};
        \node[bd, fill=white, draw=black] at (22,2.5) {};
        \node[bd, fill=white, draw=black] at (22,3.5) {};
        \node[bd, fill=white, draw=black] at (22,4.5) {};
        }
      
    \end{tikzpicture}\;
    }
    \end{center}
    \caption{On the left: usual toric diagram with the corresponding 5-brane-web at infinite coupling (in the brane-web the 7-branes where the 5-branes end are labelled by white circles). On the right: GTP with a white dot on the leftmost edge, with the corresponding 5-brane web at infinite coupling. Notice that in this case two 5-branes end on the same 7-brane.}
    \label{fig: GTP}
\end{figure}
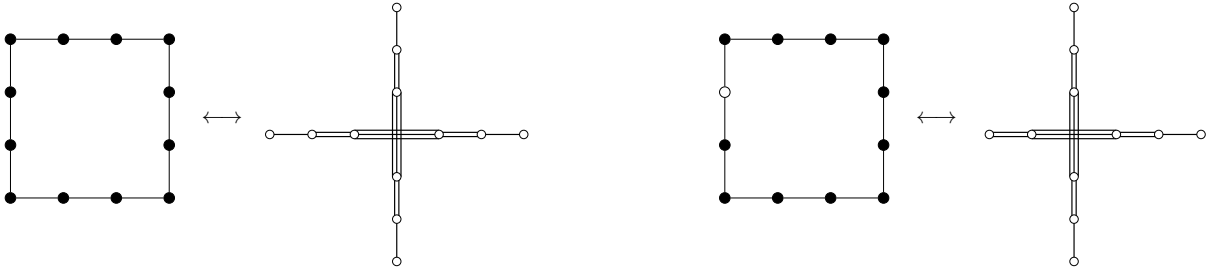

In this Section we show that our findings are compatible, and extend, results on generalized toric polygons (GTPs), which have been introduced in the seminal work \cite{Benini:2009gi}, and recently connected to the deformation theory of Calabi-Yau threefolds in \cite{Bourget:2023wlb,Alexeev:2024bko}. The cited references have explored the physical relevance of a generalization of toric geometry: ordinary toric diagrams are dual to 5-brane configurations where each 5-brane is allowed to terminate on a \textit{single} 7-brane. GTPs, instead, are dual to setups that allow for multiple 5-branes to terminate on the same 7-brane. Pictorially, this is represented by "white dots" along the external edges of the toric diagram, as opposed to the usual convention that employs "black dots" to identify non-compact divisors. See Figure \ref{fig: GTP} for an example of GTP, along with its dual brane configuration at infinite coupling. 
In turn, this breaks one of the $U(1)$ isometries of the corresponding Calabi-Yau threefold, and hence toricity. The resolution of the resulting threefold singularity can still be performed diagramatically, if one takes special care in enforcing the s-rule and the r-rule \cite{vanBeest:2020kou}.

It has been shown in \cite{Bourget:2023wlb} that, if white dots are present only on one edge of what was formerly a toric diagram, one can offer a local description of the new geometry via an auxiliary tachyon. For the full details of this construction, we refer to \cite{Bourget:2023wlb}. Here, we will only need the following facts:
\begin{itemize}
    \item an edge of lattice length $n$ corresponds to a non-compact line in the threefold supporting a singularity of type $A_{n-1}$. Adding a white dot to the edge reduces the singularity.
    \item a tachyon $T$ is a map
    \begin{equation}
        E_1 \xrightarrow[]{\hspace{0.5cm}T\hspace{0.5cm}} E_2,
    \end{equation}
    between vector bundles $E_1$ and $E_2$. Physically, the tachyon describes a Type IIA configuration with D6-branes, sourced from the original M-theory description after a suitable dimensional reduction on an $S^1$. The locus of the D6-branes is given by the locus in which the tachyon fails to be invertible.\footnote{One should more precisely think about this description in terms of tachyon condensation: the operator $T$ describes tachyonic strings stretching between a stack of D8 and anti-D8 branes, labelled by $E_1$ and $E_2$ respectively. Giving a non-trivial vev to $T$ triggers condensation between the two stacks. If the annihilation between the stacks is incomplete, there can be surviving D6-branes. This is encoded in a non-trivial profile of the tachyon. See \cite{Sen:1998sm,Collinucci:2014qfa} for additional details on this construction.}
    \item In the language of this work, one considers M-theory setups of the type exhibited in \eqref{general molecule}. As we have reviewed, one can descend to type IIA reducing along the circle contained in the $\mathbb{C}^*$-fibration $uv = \ldots$. Then the brane locus is as in \eqref{brane locus UV}. The corresponding tachyon is a matrix representing an element of $\text{Hom}(E_1,E_2)$.\footnote{In practice, we will always deal with the case of split $E_1,E_2$ bundles. Hence, we can always globally represent an element of $\text{Hom}(E_1,E_2)$ with a matrix whose entries are sections of appropriate \textit{line} bundles.} The non-compact D6 branes described by the tachyon correspond to an external edge of the toric diagram. As an example, consider the configuration on the left in Figure \ref{fig: GTP}, namely \textit{with no white dots}. This corresponds to the M-theory background:
    \begin{equation}
    \begin{cases}
        xy = z^3\\
        uv =z^3\\
        \end{cases}.
    \end{equation}
    We have seen in Section \ref{sec: HB quivers T33} that, choosing the vertical Type IIA reduction, the brane locus is:
    \begin{equation}\label{brane locus box}
        \Delta = z_1^3e_1^3e_2^3z_2^3 = 0,
    \end{equation}
    where $z_1^3$ and $z_2^3$ describe the two stacks of non-compact D6-branes on the two external edges, on the left and on the right of the toric diagram. In the threefold, these correspond to two non-compact lines supporting $A_2$ singularities. Then the data of the corresponding tachyon map is:
    \begin{equation}
        \mathcal{O}(-1,0)^{\oplus 3}\oplus \mathcal{O}(-2,1)^{\oplus 3}\oplus \mathcal{O}(-1,2)^{\oplus 3}\oplus \mathcal{O}(0,-1)^{\oplus 3} \xrightarrow[]{\hspace{0.5cm}T\hspace{0.5cm}} \mathcal{O}^{\oplus 12},
    \end{equation}
    \begin{equation}\label{tachyon 1}
        T = \left(\begin{array}{cccc}
          z_1 \mathbbm{1}_{3\times 3}  & & &  \\
             & e_1 \mathbbm{1}_{3\times 3} & & \\
             & & e_2 \mathbbm{1}_{3\times 3}& \\
            & & & z_2 \mathbbm{1}_{3\times 3}\\
        \end{array} \right).
    \end{equation}
    It is immediate to check that the tachyon \eqref{tachyon 1} is non-invertible precisely on the brane locus \eqref{brane locus box}, since its determinant is
    \begin{equation}
        \text{det}(T) = z_1^3e_1^3e_2^3z_2^3.
    \end{equation}
    \item As argued by \cite{Bourget:2023wlb}, a white dot along one edge is implemented as a nilpotent vev in the tachyon. Consider now the diagram on the right of Figure \ref{fig: GTP}, with the white dot inserted. This is described in the tachyon as:
    \begin{equation}\label{tachyon nilpotent}
        T = \left(\begin{array}{ccc|ccc}
          z_1   & \frac{1}{z_2e_1e_2}&0 & & &  \\
          0   & z_1&0 & & &  \\
          0   &0 & z_1& & &  \\
          \hline
            && & e_1 \mathbbm{1}_{3\times 3} & &  \\
            && & & e_2 \mathbbm{1}_{3\times 3}&  \\
            &&& & & z_2 \mathbbm{1}_{3\times 3} \\
        \end{array} \right),
    \end{equation}
    where we have explicitly represented the D6-brane stack labelled by $z_1$ as a 3$\times$3 matrix. Notice that the nilpotent vev in \eqref{tachyon nilpotent} does not modify its determinant, and hence the D6-brane stack. Nonetheless, the effect of the white dot on the threefold geometry can be explicitly detected via a Hanany-Witten move, that acts on the 5-branes ending on the same 7-brane: this is captured by turning on a vev along the Slodowy slice of the nilpotent part of \eqref{tachyon nilpotent}, intersected with the Levi subalgebra of the roots of the $A_{2}$ algebra labelling the nilpotent vev. All in all, this amounts to:
      \begin{equation}\label{tachyon nilpotent HW}
        T = \left(\begin{array}{ccc|ccc}
          z_1   & \frac{1}{z_2e_1e_2}&0 & & &  \\
        \alpha z_2^2e_2   & z_1&0 & & &  \\
          0   &0 & z_1& & &  \\
          \hline
            && & e_1 \mathbbm{1}_{3\times 3} & &  \\
            && & & e_2 \mathbbm{1}_{3\times 3}&  \\
            &&& & & z_2 \mathbbm{1}_{3\times 3} \\
        \end{array} \right).
    \end{equation}
    The determinant of the tachyon, and hence the brane locus, is modified as:
    \begin{equation}
        \text{det}(T) = z_1^3e_1^3e_2^3z_2^3 + \alpha z_1 e_1^2 e_2^3 z_2^4.
    \end{equation}
    Uplifting the result back in M-theory, one obtains that the presence of a white dot, combined with a Hanany-Witten move, deforms the threefold as:
    \begin{equation}\label{white dot def}
        \begin{cases}
            xy=z^3\\
            uv = z^3 +\alpha yz
        \end{cases}.
    \end{equation}
\end{itemize}
We can now come to the relation of this technology to our work on normalizable deformations of Calabi-Yau threefolds with non-isolated singularities. Notice that the deformation in \eqref{white dot def} appears among the normalizable deformations of the toric threefold, that we have listed in \eqref{def CY3 T33}. Hence our formalism naturally incorporates the effect of white dots on the starting toric geometry. 

It is crucial to remark, though, that \textit{not all deformations arise from white dots}. This is easily proven in full generality. For the sake of concreteness, consider the tachyon in \eqref{tachyon 1}. The white dots on the stack of branes $z_1^3=0$ are classified by nilpotent vevs, which are in one-to-one correspondence with ordered partitions of 3, since the stack corresponds to a $A_2$ algebra. Clearly, in \eqref{tachyon nilpotent} we have turned on a vev along the $[2,1]$ nilpotent orbit. Similarly, turning on a vev along the $[3]$ nilpotent orbit, and then performing a HW move, produces the following deformed M-theory geometry:
\begin{equation}
    \begin{cases}
        xy = z^3\\
        uv = z^3+\alpha y\\
    \end{cases},
\end{equation}
which is also a special case of \eqref{def CY3 T33}.
Finally, the orbit $[1,1,1]$ is just the undeformed tachyon. Therefore, all deformations encoded by white dots are encompassed by our analysis in the previous sections.\\

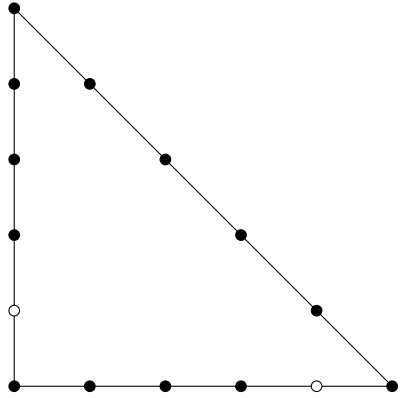
\begin{figure}[H]
\centering
    \scalebox{1.}{
    \begin{tikzpicture}
        \draw (0,0)--(5,0)--(0,5)--(0,0);
        \filldraw (0,0) circle (2pt);
        \draw[fill = white] (0,1) circle (2pt);
        \filldraw  (0,2) circle (2pt);
        \filldraw (0,3) circle (2pt);
        \filldraw (0,4) circle (2pt);
        \filldraw (0,5) circle (2pt);
        \filldraw (1,0) circle (2pt);
        \filldraw (2,0) circle (2pt);
        \filldraw (3,0) circle (2pt);
        \draw[fill = white]  (4,0) circle (2pt);
        \filldraw (5,0) circle (2pt);
        \filldraw (4,1) circle (2pt);
        \filldraw (3,2) circle (2pt);
        \filldraw (2,3) circle (2pt);
        \filldraw (1,4) circle (2pt);
    \end{tikzpicture}}
    \caption{GTP for the $T_5$ theory with two white dots.}
    \label{fig:T5 white dots}
    \end{figure}

Moreover, one can think of turning on white dots on multiple non-parallel edges of the toric diagram on the left of Figure \ref{fig: GTP}. In such case, the Type IIA description is unavailable, and with it goes the possibility to capture the geometry via a tachyon description. Nonetheless, the deformed threefolds related to white dots on multiple edges of the toric diagrams can be predicted with the powerful new techniques based on \textit{decorated toric polygons} and \textit{scattering diagrams}, due to \cite{Alexeev:2024bko}. E.g.\ in Example 6.15 of \cite{Alexeev:2024bko}, the GTP in Figure \ref{fig:T5 white dots} is considered.

It is a GTP for the $T_5$ theory with white dots on two non-parallel edges: hence no tachyon describing this configuration exists. It is then proven that the associated deformed geometry is:
\begin{equation}\label{Bousseau def}
    xyz = w^5 + c_1 x w^3 +c_2 y w^3. 
\end{equation}
It is straightforward to check that the deformations in \eqref{Bousseau def} are dynamical, and are a special case of the fully deformed $T_5$ threefold, shown in \eqref{Tn hyp fully def}. As a concluding remark, notice that the deformations corresponding to white dots on multiple sides \textit{do not exhaust all dynamical deformations}. For example, the dynamical deformations in the term $\sum\limits_{i=0}^{n-1}\overline{c}_i w^i$ of \eqref{Tn hyp fully def} can clearly not be realized via the white dot technology. The rationale from the 5-brane perspective is that these deformations correspond to movements, in the plane orthogonal to the brane-web, of 5-brane subwebs that stretch along multiple edges of the corresponding toric diagrams. This is opposed to subwebs removed by white dots, which for the box diagrams and $T_n$ theories affect each edge independently (namely, a white dot on a given edge only removes subwebs of 5-branes related to that edge).

\section*{Acknowledgements}
We would like to thank Antoine Bourget, Cyril Closset, Andr\'es Collinucci, Sara Pasquetti, Palash Singh, Sebastiano Garavaglia,  Sakura Sch\"afer-Nameki for discussions. The research of AS is funded from the VR Centre for Geometry and Physics (VR grant No.\ 2022-06593) at Uppsala University. AS and MDZ acknowledge the European Union for funding (ERC, HIGH, 101171852). Views and opinions expressed are however those of the authors only and do not necessarily reflect those of the European Union or the European Research Council. Neither the European Union nor the granting authority can be held responsible for them. AS and MDZ also acknowledge the VR project grant No. 2023-05590 for partial support.
The research of M.D.M. is also funded through an ARC advanced project, and further supported by IISN-Belgium (convention 4.4503.15).
JFG is supported by the EPSRC Open Fellowship (Schafer-Nameki) EP/X01276X/1 and the ``Simons Collaboration on Special Holonomy in Geometry, Analysis and Physics''. JFG is grateful for the hospitality of the Centre for Geometry and Physics at Uppsala University, where part of this project was completed. The research of AS was funded by
the VR Centre for Geometry and Physics (VR grant No.\ 2022-06593) at Uppsala University.

\appendix

\section{More general toric complete intersections}\label{sec:complete intersections}
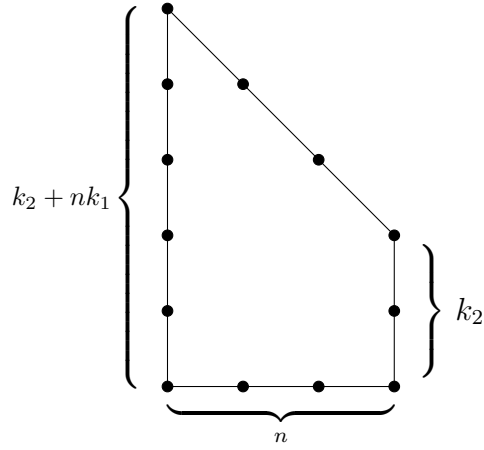
\begin{figure}[H]
\centering
    \scalebox{1.}{
    \begin{tikzpicture}
        \draw (0,0)--(3,0)--(3,2)--(0,5)--(0,0);
        \filldraw (0,0) circle (2pt);
        \filldraw (0,1) circle (2pt);
        \filldraw  (0,2) circle (2pt);
        \filldraw (0,3) circle (2pt);
        \filldraw (0,4) circle (2pt);
        \filldraw (0,5) circle (2pt);
        \filldraw (1,0) circle (2pt);
        \filldraw (2,0) circle (2pt);
        \filldraw (3,0) circle (2pt);
        \filldraw (3,1) circle (2pt);
        \filldraw (3,2) circle (2pt);
        \filldraw (2,3) circle (2pt);
        \filldraw (1,4) circle (2pt);
        \node at (1.5,-0.5) {$\underbrace{\hspace{3cm}}_{n}$};
        \node[font=\footnotesize] at (-1.4,2.5) {$k_2+nk_1$};
        \node at (-0.3,2.5) {$\begin{cases}
            \\
            \\
            \\
            \\
            \\
            \\
            \\
        \end{cases}$};
        \node at (3.3,1) {$\left.
\begin{array}{l}
\\
\\
\\
\end{array}
\right\}$};
\node at (4,1) {$k_2$};
        \end{tikzpicture}
        }
    \caption{GTP for the $T_5$ theory with two white dots.}
    \label{fig: trapezoid}
    \end{figure}

In this Section we briefly outline how our characterization of 5d Higgs Branches in terms of explicit dynamical deformations in toric Calabi-Yau threefolds can be generalized to cases beyond the box diagram and $T_n$ theories examined in the previous Sections.\\
\indent For instance, a natural extension of our framework also applies also to the complete intersections:
\begin{equation}\label{eq trapezoid}
    \begin{cases}
        xy = z^n\\
        uv = y^{k_1}z^{k_2}\\
    \end{cases}.
\end{equation}
These are toric diagrams which have the shape of a trapezoid, as shown in Figure \ref{fig: trapezoid}.
Let us assume $k_2>0$ for simplicity. The case $k_2=0$ can be treated analogously.

\begin{figure}[H]
\begin{center}
 \scalebox{1}{\begin{tikzpicture}
            \node[gauge,label=below:{$4$}] (c1) at (0,0) {};
            \node[gauge,label=below:{$3$}] (t1) at (1,0) {};
            \node[gauge,label=below:{$2$}] (t2) at (2,0) {};
            \node[gauge,label=below:{$1$}] (t3) at (3,0) {};
            \node[gauge,label=below:{$2$}] (q1) at (-1,0) {};
            \node[gauge,label=below:{$1$}] (q2) at (-2,0) {};
            \node[gauge,label=above:{$3$}] (r1) at (0,1) {};
            \node[gauge,label=above:{$2$}] (s1) at (1,1) {};
            \node[gauge,label=above:{$1$}] (s2) at (2,1) {};
            \node[gauge,label=above:{$2$}] (s3) at (-1,1) {};
            \node[gauge,label=above:{$1$}] (s4) at (-2,1) {};
            \draw (c1)--(t1)--(t2)--(t3) (c1)--(q1)--(q2) (c1)--(r1)--(s1)--(s2) (r1)--(s3)--(s4);
    \end{tikzpicture}}
    \end{center}
     \caption{Magnetic quiver for the geometry in \eqref{eq trapezoid} with $n=3,k_1=1,k_2=2$.}
    \label{fig: trapezoid MQ}
\end{figure}
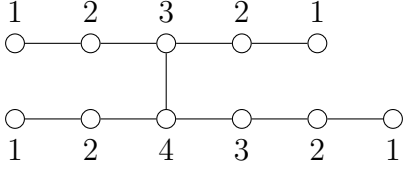

The dynamical deformations of \eqref{eq trapezoid} can be readily computed as:
\begin{equation}\label{eq trapezoid deformed}
    \begin{cases}
        xy = z^n  + \sum\limits_{\substack{i\ge 1,j \ge 0 \\ i+j \le n-1}}c_{ij}u^{i}z^j+ \sum\limits_{\substack{i\ge 1,j \ge 0 \\ i+j \le n-1}}\hat{c}_{ij}v^{i}z^j + \sum\limits_{\substack{i\ge 1}}\tilde{c}_{i}z^i,\\
        uv = y^{k_1}z^{k_2} + \sum\limits_{\substack{0< i \le k_2\\0\le j\le k_2-i}}d_{ij}y^{i}z^j + \sum\limits_{\substack{0\le i < n k_1+k_2-1\\0\le j<n k_1+k_2-1-i}}\hat{d}_{ij}x^{i}z^j. \\
    \end{cases}
\end{equation}
E.g.\ for $n=3,k_1=1,k_2=2$ the fully deformed threefold reads:
\begin{equation}\label{trapezoid example}
\begin{cases}
       xy = z^3+c_1u+c_2u z+c_3u^2+c_4v+c_5vz+c_6v^2+c_7 z+c_8 z^2, \\
       \begin{split}
       uv =& \hspace{0.1cm} yz^2+c_9 y+c_{10} yz+c_{11}y^2+c_{12}+c_{13}z+c_{14}z^2+c_{15}z^3+c_{16}x+\\
    &+c_{17}xz+c_{18}xz^2+c_{19}x^2+c_{20}x^2z+c_{21}x^3.
       \end{split}\\
       \end{cases}
\end{equation}
The magnetic quiver can be readily extracted from the brane web, and it is depicted in Figure \ref{fig: trapezoid MQ}.

The dimension of the MQ Coulomb Branch is 21, in agreement with the amount of normalizable deformations in \eqref{trapezoid example}. Notice that one could also have interpreted the undeformed geometry corresponding to \eqref{trapezoid example} as a leaf in the Hasse diagram of the 5d SCFT engineered by \eqref{box diagram eq}, for some choice of $n$ and $k$. This is evident from the magnetic quiver perspective. Naturally, the deformation theory of \eqref{trapezoid example} can be retrieved also from this alternative picture. For generic $n$,$k_1$,$k_2$ the total number of dynamical deformations is:
\begin{eqnarray}
    \# \text{of dynamical deformations}_{n,k_1,k_2} =\scalemath{1}{ (n-1)(n+1)
+ \frac{(k_2+n k_1-1) (k_2+n
   k_1)}{2}+\frac{k_2 (k_2+1)}{2}}. \nonumber \\
\end{eqnarray}

\section{Notation for Quivers and Brane Webs}
\label{app:QuiverBWNotation}
In this appendix we briefly summarise the notation we use for quivers and brane webs.

\subsection{Quivers}
\label{app:QuiverNotation}
As is customary, we use square nodes for flavour groups, round nodes for gauge groups, and edges for bifundamental hypermultiplets. We colour the nodes white for unitary groups and yellow for special unitary groups (possibly with Chern-Simons level $k$)
:
\begin{equation}
    \begin{tikzpicture}
        \node[gauge,label=below:{$n$}] at (0,0) {};
        \node at (1,0) {$=$};
        \node[gauge,label=below:{U($n$)}] at (2,0) {};
        \node at (2.8,0) {,};
        \node[gaugey,label=below:{$(n)_k$}] at (4,0) {};
        \node at (5,0) {$=$};
        \node[gaugey,label=below:{SU$(n)_k$}] at (6,0) {};
        \node at (6.8,0) {,};
        \node[gaugey,label=below:{$n$}] at (8,0) {};
        \node at (9,0) {$=$};
        \node[gaugey,label=below:{SU$(n)_0$}] at (10,0) {};
    \end{tikzpicture}\;.
\end{equation}

\paragraph{Unframed Quivers.}
If the quiver is unframed, i.e.\ there are no flavour nodes, then there can be a 1-form symmetry which can be gauged. (see e.g.\ \cite{Bourget:2020xdz}, where the 1-form symmetry is referred to as '$H$'.).

For an unframed quiver with only unitary gauge nodes there is a U$(1)^{(1)}$ 1-form symmetry which we always take to be gauged, and hence the true gauge group of the quiver is $\prod_i\mathrm{U}(r_i)/\mathrm{U}(1)$, where $r_i$ are the ranks of the individual gauge nodes.

\subsection{Brane Webs}
\label{app:BWNotation}
We consider webs of $(p,q)5$-branes and $[p,q]7$-branes in Type IIB String Theory \cite{Aharony:1997ju,Aharony:1997bh,DeWolfe:1999hj}. The occupied spacetime directions of the branes are summarised in Table \ref{tab:spacetime}.

\begin{table}[h]
\begin{center}
\begin{tabular}{|c|c|c|c|c|c|c|c|c|c|c|}
\hline
Type IIB & $x^0$ & $x^1$ & $x^2$ & $x^3$ & $x^4$ & $x^5$ & $x^6$ & $x^7$ & $x^8$ & $x^9$\\
\hline
$(p,q)5$-brane & $\times$ & $\times$ & $\times$ & $\times$ & $\times$ & \multicolumn{2}{c|}{angle $\alpha$} & & & \\
\hline
$[p,q]7$-brane & $\times$ & $\times$ & $\times$ & $\times$ & $\times$ & & & $\times$ & $\times$ & $\times$ \\ \hline
\end{tabular}
\caption{Occupation of space-time directions of the $(p,q)5$-branes and $[p,q]7$-branes in Type IIB are denoted by $\times$. The angle $\alpha$ depends on the $(p,q)$ charges and the axio-dilaton $\tau$; $\alpha=\arg(p+\tau q)$. We set $\tau=i$ in the rest of the paper, s.t.\ $\tan(\alpha)=q/p$. Our 5d $\mathcal{N}=1$ theories exist as effective field theories living on fivebranes suspended between sevenbranes.}
\label{tab:spacetime}
\end{center}
\end{table}

We depict brane webs by drawing the $(x^5,x^6)$ plane:
\begin{equation}
    \begin{tikzpicture}
        \node[seven,label=above:{\scriptsize$[1,1]$}] (1) at (2,2) {};
        \node[seven,label=below:{\scriptsize$[1,0]$}] (2) at (-2,0) {};
        \node[seven,label=below:{\scriptsize$[0,1]$}] (3) at (0,-2) {};
        \draw (2)--(0,0)--(3) (0,0)--(1);
        \draw[dash dot] (1)--(4,2) (2)--(-4,0) (3)--(2,-2);
        \node at (-1,0.3) {\scriptsize$(1,0)$};
        \node at (0.7,-1) {\scriptsize$(0,1)$};
        \node at (1.5,0.75) {\scriptsize$(1,1)$};

        \draw[->] (-6,-2)--(-4,-2);
        \draw[->] (-6,-2)--(-6,0);
        \node at (-6,0.3) {$x^6$};
        \node at (-3.5,-2) {$x^5$};
    \end{tikzpicture}\;.
\end{equation}
A $(p,q)5$-brane ends on a $[p,q]7$-brane, or on a fivebrane vertex which preserves $(p,q)$ charges. Multiple $(p,q)5$-branes can end on the same $[p,q]7$-brane as long as the s-rule \cite{Hanany:1996ie,Benini:2009gi,vanBeest:2020kou,Bergman:2020myx} is not violated.

Each $[p,q]7$-brane induces an SL$(2,\mathbb{Z})$ monodromy cut (depicted by a dot-dash line) with associated monodromy matrix
\begin{equation}
    M_{[p,q]}=\begin{pmatrix}
                1-pq & p^2\\
                -q^2 & 1+pq
            \end{pmatrix}\;.
\end{equation}
We usually don't draw the monodromy cuts of sevenbranes if they are oriented in such a way that they do not cross any fivebranes.

An $(r,s)5$-brane crossing the monodromy cut of a $[p,q]7$-brane is affected in the following way:
\begin{equation}
    \begin{tikzpicture}
            \draw (0,0)--(2,0)--(2,-2);
            \node[seven,label=below:{\scriptsize$[p,q]$}] (1) at (1,-1) {};
            \draw[dash dot] (1)--(3,1);
            \node at (1,0.5) {\scriptsize$(r,s)$};
            \node at (3,-1) {\scriptsize$M_{[p,q]}.(r,s)$};
            \node at (4,1.3) {\scriptsize$[p,q]$ monodromy cut};
            \draw[dash dot] (3.6,1.6)--(4,2);
        \end{tikzpicture}\;.
\end{equation}
When the $(r,s)5$-brane is pulled through the $[p,q]7$-brane, $|ps-qr|$ extra $(r,s)5$-branes are created via the Hanany-Witten effect:
\begin{equation}
    \begin{tikzpicture}
            \draw (0,0)--(2,0)--(2,-2);
            \node[seven,label=right:{\scriptsize$[p,q]$}] (1) at (4,2) {};
            \node at (1,0.5) {\scriptsize$(r,s)$};
            \node at (3,-1) {\scriptsize$M_{[p,q]}.(r,s)$};
            \draw[thick,double] (2,0)--(1);
            \node at (4,1) {\scriptsize$|ps-qr|(p,q)$};
        \end{tikzpicture}\;.
\end{equation}

\bibliographystyle{at}
\bibliography{bibliography.bib}
\end{document}